\documentclass[twocolumn]{aastex63}
\usepackage[utf8]{inputenc}
\usepackage{amsthm}
\usepackage{amsmath}
\usepackage{hyperref}
\usepackage{booktabs}
\usepackage{wrapfig}
\usepackage{wasysym}
\usepackage{float}
\usepackage{xcolor}

\usepackage{graphicx}
\usepackage{url}
\graphicspath{{./}{figures/}}
\usepackage{natbib}
\listofchanges
\usepackage{comment}
\shorttitle{Star Formation Histories from SEDs and CMDs}
\shortauthors{Olsen et al.}
\begin{document}
\received{March 5, 2021}
\submitjournal{ApJ}
\accepted{March 29, 2021}
\title{
Star Formation Histories from SEDs and CMDs Agree:  Evidence for Synchronized Star Formation in Local Volume Dwarf Galaxies over the Past 3 Gyr

}

\correspondingauthor{Charlotte Olsen}
\email{olsen@physics.rutgers.edu}

\author[0000-0002-8085-7578]{Charlotte Olsen}
\affiliation{ 
Department of Physics and Astronomy, Rutgers University,  
Piscataway, NJ 08854, USA}

\author[0000-0003-1530-8713]{Eric Gawiser}
\affiliation{ 
Department of Physics and Astronomy, Rutgers University,  
Piscataway, NJ 08854, USA}

\author[0000-0001-9298-3523]{Kartheik Iyer}
\affiliation{Dunlap Institute, University of Toronto, Toronto, ON, Canada}

\author[0000-0001-5538-2614]{Kristen B. W.  McQuinn}
\affiliation{
Department of Physics and Astronomy, Rutgers University,  
Piscataway, NJ 08854, USA} 

\author[0000-0002-9280-7594]{Benjamin~D.~Johnson}
\affiliation{Center for Astrophysics $|$ Harvard \& Smithsonian, 60 Garden Street, Cambridge, MA 02138, USA}

\author[0000-0003-4122-7749]{Grace Telford}
\affiliation{
Department of Physics and Astronomy, Rutgers University,   
Piscataway, NJ 08854, USA}

\author[0000-0002-1685-5818]{Anna C. Wright}
\affiliation{Department of Physics \& Astronomy, Johns Hopkins University, 3400 N. Charles Street, Baltimore, MD 21218, USA}

\author[0000-0002-7767-5044]{Adam Broussard}
\affiliation{
Department of Physics and Astronomy, Rutgers University, 
Piscataway, NJ 08854, USA}

\author[0000-0002-8816-5146]{Peter Kurczynski}
\affiliation{Goddard Space Flight Center, Code 665, Greenbelt, MD 20771, USA}

\begin{abstract}
     Star Formation Histories (SFHs) reveal physical processes that influence how galaxies form their stellar mass. We compare the SFHs of a sample of 36 nearby (D $\lessapprox$ 4 Mpc) dwarf galaxies from the ACS Nearby Galaxy Survey Treasury (ANGST), inferred from the Color Magnitude Diagrams (CMDs) of individually resolved stars in these galaxies, with those reconstructed by broad-band Spectral Energy Distribution (SED) fitting using the Dense Basis SED fitting code.
     When comparing individual SFHs, we introduce metrics for evaluating SFH reconstruction techniques. 
     For both the SED and CMD methods, the median normalized SFH of galaxies in the sample shows 
     a period of quiescence at lookback times of 3-6 Gyr followed by rejuvenated star formation over the past 3 Gyr that remains active until the present day.   To determine if these represent special epochs of star formation in the D $<$ 4~Mpc portion of the 
     Local Volume, we break this ANGST dwarf galaxy sample into subsets based on specific star formation rate and spatial location.  Modulo offsets between the methods of about 1 Gyr, all subsets show significant decreases and increases in their median normalized SFHs at the same epochs, and the majority of the individual galaxy SFHs are consistent with these trends.    
     These results motivate further study of potential synchronized star formation quiescence and rejuvenation in the Local Volume as well as development of a hybrid method of SFH reconstruction that combines CMDs and SEDs, which have complementary systematics.

\end{abstract}

\section{Introduction}
A fuller understanding of galaxy evolution 
requires
disentangling the many physical processes acting within and upon galaxies over cosmic time. One of the most fundamental of these processes is star formation. In situ star formation can be either enhanced or suppressed by external environmental factors such as major and minor mergers, tidal stripping, or gas enrichment from tidal streams. This is reinforced by observations that show  properties that correlate with galaxy environment include color, morphology, gas content, and star formation rate (SFR). Co-evolution of properties of galaxies sharing the same environment is referred to as galactic conformity \citep[e.g.][]{Kauffmann2012:1209.3306v2,Hearin2015:1504.05578v1, Sin2017:1702.08460v2, Calderon2017:1712.02797v2, Rafieferantsoa2017:1707.01950v2,Sin2019:1902.09543v1}.  Galactic conformity falls into two categories stemming largely from the work in \cite{Kauffmann2012:1209.3306v2}. These categories are one-halo conformity, where the galaxies in question share the same dark matter halo, and two-halo conformity, where the galaxies span a greater distance. \cite{Kauffmann2012:1209.3306v2} found evidence for two-halo conformity spanning up to 4~Mpc separations, but \cite{Sin2017:1702.08460v2} demonstrated that this could be a selection effect. 
 Galactic conformity has also been studied in other properties, including morphology \citep{Calderon2017:1712.02797v2, Otter2020:2001.01231v1}, color \citep{Bray2015:1508.05393v2,Calderon2017:1712.02797v2, Treyer2017:1712.05318v2}, and specific star formation rate (sSFR) \citep{Rafieferantsoa2017:1707.01950v2, Calderon2017:1712.02797v2}. Recent methodological improvements in the reconstruction of galaxy  star formation histories (SFHs) have created an opportunity to look at conformity as a function of time, which might identify epochs of synchronized star formation or quiescence. 

Dwarf galaxies exist in a wide variety of environments, from isolated field galaxies, to large groups of satellites. Due to their low masses, dwarf galaxies are particularly sensitive to the effects of environment \citep{Geha2006, Geha2012, Weisz2014reion, Weisz2015}. The diverse star formation histories of dwarf galaxies help to illustrate  the connection between star formation mechanisms and environment\citep[e.g.][]{Tolstoy2009,McQuinn2011, weisz, Gallart2015, Cignoni2019}. Dwarfs are believed to carry the imprint of reionization \citep[e.g.][]{Tolstoy2009,weisz2014UV, Weisz2014reion}, and provide analogues for young galaxies observed at high redshift \citep{Ouchi2020}. Dwarf galaxies are also the most numerous galaxies in the universe. 


Nearby galaxies provide ideal laboratories for deepening our knowledge of star formation, as individual stellar populations can be resolved. Simulated populations from stellar population synthesis (SPS) modeling can be compared directly with observations and  compared to isochrones of known age to determine when stars formed. Early efforts required generating large numbers of synthetic color magnitude diagrams (CMDs) with different combinations of stellar evolutionary tracks and therefore each with a known SFH, and then matching the observed CMD to the closest simulated CMD to determine the SFH \citep{Faber,Tosi1989,Bertelli1992}. Without placing constraints on unknown parameters within the model, this method is either too computationally costly, or largely qualitative. These drawbacks are avoided by binning in the CMD and then performing a $\chi^2$ minimization over each time bin  \citep{Dolphin1997,Aparicio1997}.    
This method requires careful bookkeeping 
of the probability that a star or `point' falls within a particular bin, since decreasing the number of points in one bin will increase the number of points in the adjacent bins. Bin size can be chosen to match the scale of the smallest robust features in the CMD. When binning stars by age, SFHs are recovered as flat star formation rates for each time bin. A detailed account of how binning is currently handled in CMDs can be found in \cite{dolphin}. 
CMD based star formation histories can recover major features (e.g., bursts or quiescent episodes) in galaxy SFHs and can deliver unparalleled time resolution for recent star formation \citep[e.g.,][]{ weisz,McQuinn2011, johnson, mcquinnUV}. The dominant systematics in the CMD method come from 
the model assumptions in stellar evolution libraries even in deep observations \citep{Aparicio2009,weisz, Dolphin2012}.   

When CMDs are not available, SFHs can be inferred via Spectral Energy Distribution (SED) fitting techniques \citep[e.g.][]{Rix'n'Rieke1993, Sawicki1998, Bell'n'deJong2000,Brinchmann2000, Shapley2001, Bell'n'deJong2001, Papovich2001} that take advantage of the information encoded over a wide range of wavelengths. The basic premise is that by fitting either observed galaxy spectra or photometry to a modeled spectrum, one can decompose the properties and components of that spectrum. 
SED fitting seeks to measure the SFR and stellar mass (M$_*$) while also accounting for other parameters that may be degenerate with these including metallicity (Z), dust attenuation ($A_V$), and the shape of the SFH. Some early SED fitting methods recovered SFHs from spectra via  a fairly straightforward inversion method  \citep{Heavens2000, Fernandes2005,Tojeiro2007}, while still managing to recover galaxy properties well, but SFHs can be poorly constrained \citep{Ocvirk2006}. 
SED fitting codes designed to work with broad-band photometry typically model the SFH as a simple parametric form such as constant, exponentially declining, linear-exponential \citep{Lee2014}, lognormal, or double power laws \citep[e.g.][]{Buat2008, Maraston2010, Papovich2011, Diemer2017}; however, these simple models cannot reveal complexities in the SFHs resulting from the physical processes driving star formation. To better represent such features, non-parametric codes that allow for more flexible SFHs \citep[e.g.][]{CidFernandes2005, Ocvirk2006, Leja2017, iyer2017,Iyer2019,Robotham2020,Johnson2020} recover  meaningful features in galaxy SFHs while reducing the bias in recovered galaxy properties. 

SED fitting is affected by different systematics than the CMD method. SEDs offer more information about numerous, low-luminosity, older-lived stars via observations that span redder wavelengths where these stars dominate the integrated light. However, it can be a challenge to disentangle degeneracies between dust, metallicity, redshift, and nebular and dust emission to determine the contribution of each to the integrated light \citep{walcher, conroyrev}.

 While SED fitting began as a solution for the high redshift universe, the ability of the non-parametric codes to recover properties in low redshift galaxies presents  several new opportunities. Together CMD and SED methods allow for decomposition of observational systematics and methodological differences as well as combining information from the CMD's resolution with the SED's broad wavelength range. We can compare the results of SED- and CMD-based SFH estimation on the same set of galaxies to 1) calibrate our ability to recover galaxy properties against the current gold standard of the CMD method, 2) investigate if there is additional information gained through either method that cannot be recovered from the other, 3) find insights revealed through agreements between the two methods since good agreement would imply a robust result, and 4) consider  ways to combine the strengths of these methods due to their complementary systematics. 

In this paper,  Section \ref{sec:Data} introduces the data set,  including the broadband photometry  and CMD SFHs. Section \ref{sec:method} describes our methodology, including generation and validation of mock photometry before the process of SFH reconstruction. Results for individual galaxies are given in Section \ref{sec:individual}, the unexpected agreement of median normalized SFHs is described in Section \ref{sec:median}, and we discuss our conclusions in Section \ref{sec:Disc}. 
In calculating the age of the universe, we assumed a $\Lambda$CDM cosmology with $H_0=70\textrm{ km s}^{-1}\textrm{Mpc}^{-1}$ and $\Omega_m = 0.3$.  Since the broadband photometry was provided in AB absolute magnitudes, our choice of cosmology would not affect the results of SED fitting, where for consistency all galaxies were assigned a  redshift of $2.3 \times 10^{-9}$ to correspond to 10 pc.  


\section{The Data}\label{sec:Data}

    We have selected a sample of galaxies from the ACS Nearby Galaxy Survey Treasury (ANGST), a volume-limited survey of 69 nearby galaxies ($D<4$ Mpc) with multi-wavelength photometry of resolved stars. The ANGST sample offers a set of  of primarily low mass galaxies with masses $<10^9M_\odot$. These galaxies have diverse morphologies in a wide variety of environments from close groups to isolated systems \citep{dalcanton}. 
    Within the ANGST sample, there are 60 dwarf galaxies, and 50 of those whose broadband multi-wavelength (UV through NIR) photometry matches the HST footprint, enabling a meaningful comparison between the resolved and integrated photometry \citep{johnson}.
      The individual star observations used to create CMDs, from which SFHs were derived in \cite{weisz}, are available publicly through the Barbara A. Mikulski Archive for space telescopes (MAST).

\subsection{Broadband Photometry} \label{sec:phot}
We utilize the photometry made publicly available in \cite{johnson}, which offers a  thorough description of the observations. FUV and NUV band observations came from The Galaxy Evolution Explorer (GALEX). For galaxies within the Sloan Digital Sky Survey (SDSS) footprint, the u,g,r,i,z bands were used for optical imaging. Additional ground-based optical observations were taken in the Johnson-Cousins UVBR filters to match SDSS observations \citep{cook} for galaxies in the sample that fell outside the survey footprint of SDSS. IR photometry came from Spitzer Infrared Array Camera (IRAC) obtained from the Local Volume Legacy (LVL) and 11 Mpc H$\alpha$ and UV Galaxy Survey (11HUGS) surveys \citep{Lee2009,Dale2009,johnson,Cook2014}. 
Galaxy SFHs cannot be  constrained well with SEDs that include only a few bands or that do not sample the range of UV through near-IR as discussed in the reviews by  \citet{walcher} and \citet{conroyrev}. For this reason we limit our sample to galaxies from \cite{johnson} for which we have integrated photometry in at least six bands, of which at least one is in the UV, and which include both the 3.6 and 4.5$\mu$m IRAC bands. Note that this does not formally require a photometric detection in any particular band.  This leaves us with a sample of 36 galaxies.
Uncertainties were not publicly available for the broadband photometry, so we conservatively assumed 10\% uncertainties on all bands.     

\subsection{Color Magnitude Diagram Star Formation Histories}
\label{section:CMD data} 
The ANGST galaxies have a wide range of morphologies and diverse SFHs as reconstructed through their CMDs. A discussion on the CMD SFHs for the ANGST dwarf galaxies can be found in \cite{weisz, weisz2}. 
We obtained the SFHs used in the analysis of  \cite{johnson} (D. Weisz, private communication).
These SFHs have 36 time bins, each with a width of 0.1 in log time with the exception of the last, oldest bin, which covers 12.6 to 14.125 Gyrs ago. This allows for exceptional time resolution for recent star formation, but SFRs in the older bins are highly uncertain if observations do not reach the depth of the oldest main-sequence turn off \citep{WeiszlgI2014}, and this runs the risk of over-representing the time resolution in older bins.

To address this, we instead adopted the time binning for ANGST CMD SFHs used by \cite{McQuinn2010}. This coarser binning has broader logarithmic bins  during both recent epochs and at the oldest times, with bins of 0.1 dex during intermediate times, which allowed us to maintain most of the time resolution for the recent SFH without overestimating the robustness of older SFH bins.  \par

\section{Methodology} 
\label{sec:method} 


\subsection{Validation with Mock Photometry}
\label{section:phot_valid}
Comparison between methods typically used at high and low redshifts can be challenging for multiple reasons. SED fitting that uses broadband photometry typically does so in the absence of spatially resolved photometry, while the CMD method must still handle similar systematics as SED fitting despite the resolution advantage. We considered it important to try to differentiate between systematics from the observations and those due to methodological differences.

To compare the CMD SFHs with the broadband SEDs, we generated mock photometry from the CMD SFHs as in \cite{johnson}. CMD SFHs were input to the stellar population synthesis code Flexible Stellar Population Synthesis (FSPS) \cite{Conroy2009, conroygunn}, where we kept all assumptions as close to those used in \cite{johnson} as possible.
We used Padova isochrones \citep{Bertelli1994, Girardi2004, Marigo} and the MILES spectral library \citep{Sanchez-Blazquez2006} to match assumptions used in the generation of the CMD SFHs. 
A Table 
in Appendix \ref{appendix:systematics} lists the custom FSPS parameters for the ANGST sample used in \cite{johnson} as well as the ``default'' FSPS parameters in Dense Basis before customization.  We used a Salpeter IMF \citep{Salpeter1955} and a modeled as a modified power law dust law with a steep attenuation curve as used in \citet{johnson}. Metallicity estimates were set by metallicities recovered from the CMD SFH method as reported in \cite{johnson}.   

Matching the model assumptions allowed us to produce model spectra for each galaxy using its CMD SFH as an input to FSPS. We then converted these spectra into the Mock Photometry shown in the left panel of Figure \ref{fig:gp_vs_cmd}; a second mock SED that adds the near-infrared \textit{JHK} bands is termed Mock Photometry with JHK. Comparing the mock photometry to the integrated broadband photometry, which we refer to as Observed Photometry, provides a useful consistency test as the mock photometry coming from the CMD SFHs are independent of the broadband SEDs. We found overall offsets of no more than a factor of two between the modeled and observed photometry, in line with the results of \cite{johnson}. 

However, we found typically larger  offsets between mock and observed photometry in IRAC Ch3 (5.8$\mu$m) and Ch4 (8.0$\mu$m).  This is likely caused by the complexity of modeling the polycyclic aromatic hydrocarbons (PAHs) to which these bands are sensitive at low redshift; while FSPS offers PAH parameters that could be varied, this means that further information is offered by these bands, so we instead chose to exclude them from SED fitting and to turn off PAH emission in our FSPS runs.  

In addition to gauging how well the two sets of observations agreed, the generation of mock photometry allowed us to create a best-case scenario to test Dense Basis SFH reconstruction against the CMD method. Mock photometry provides the best translation of the information in the CMD SFH into photometry ready for SED-fitting. This way we could differentiate between methodological differences and observational systematics, as disagreements between the CMD SFH and the Dense Basis results using mock photometry demonstrate the former while disagreements with the Dense Basis results using observed photometry can be caused by either or both.

\subsection{Nonparametric Star Formation History Reconstruction}
\label{section:SFH_recon}
In order to reconstruct the SFHs from the broadband photometry for our galaxies, we used the state-of-the-art Dense Basis SED fitting code \citep[][see also \citealt{iyer2017}]{Iyer2019}. Dense Basis is a fully nonparametric SFH reconstruction code that uses Gaussian Processes to explore the entirety of functional space to recover the most likely SFH for the available photometry along with robust uncertainties \citep{Iyer2019}. The code is non-parametric as it avoids requiring any particular parametric form for the SFH. A majority of SED fitting codes assume parametric SFHs, either a simple functional form or SFRs binned by time much like the CMD method;  the simple functional forms include constant, exponentially declining,  double power law, and lognormal SFHs \citep{iyer2017, Carnall2019}. A drawback of parametric models of SFHs is that the priors imposed by the models can bias the resulting inferred stellar population such that too much stellar mass is allocated to younger stars \citep{iyer2017, Leja2019, Carnall2019, Lower2020}. Promisingly, nonparametric SED fitting codes result in more flexible SFH reconstruction with less biased results when recovering galaxy properties than with parametric SED fitting routines \citep{CidFernandes2005,Tojeiro2007,iyer2017,Iyer2019,Leja2019}. Nonparametric SFH models offer flexibility comparable to the CMD method.

  \begin{figure*} 
        \centering\includegraphics[width=1\textwidth]{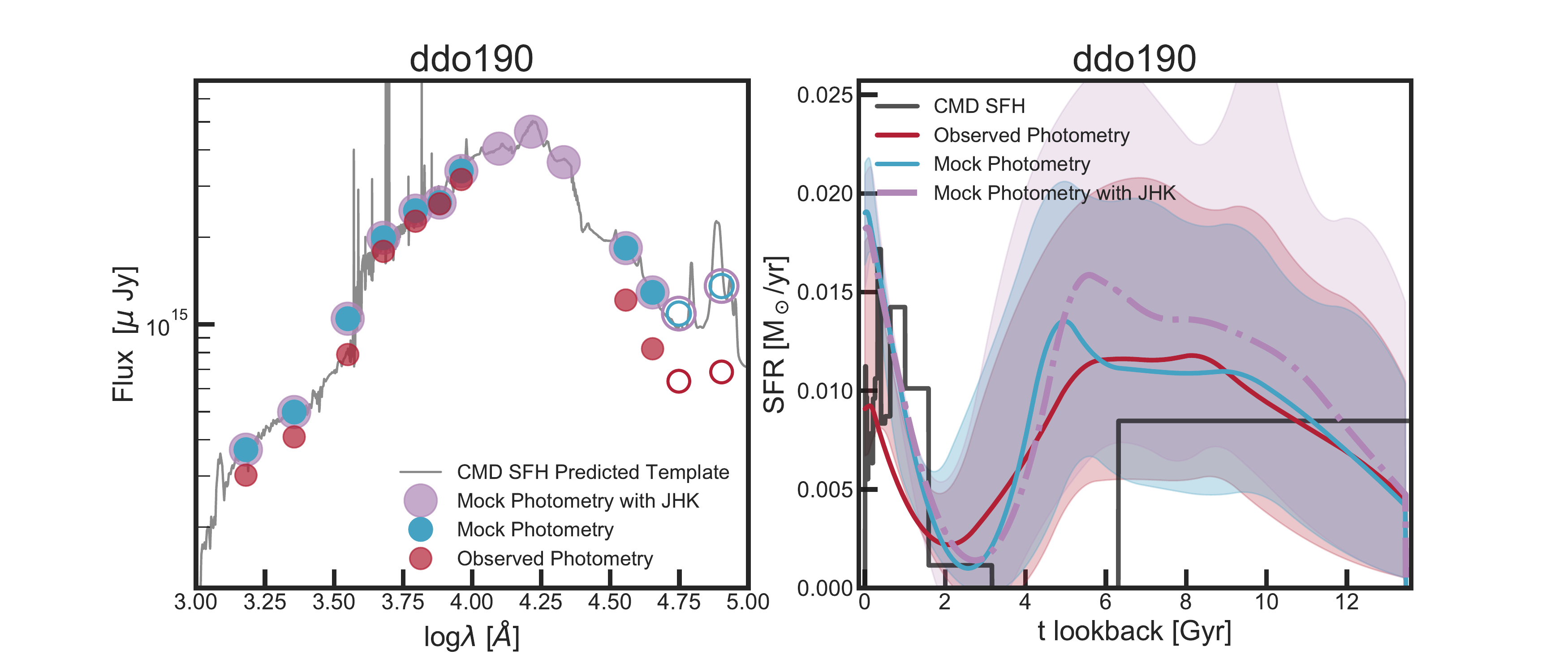}
        \caption{Example SEDs (left) and accompanying SFHs (right) for observed photometry (red) along with mock photometry generated from the CMD SFH (blue). To account for additional information about old stellar mass that might be contained in the CMDs, we create mock photometry including \textit{JHK} bands (purple) to sample around the 1.6$\mu m$ bump.  IRAC Ch3 and Ch4, which were excluded from SED fitting as described in 
        S~\ref{sec:method}, are indicated by open circles in the SEDs.  The right panel shows the SFHs and uncertainty ranges generated by Dense Basis by SED fitting for each set of photometry, with the CMD method  SFH in black for comparison.}
    \end{figure*}
    \label{fig:gp_vs_cmd}

Dense Basis strives to maximize the amount of information about the SFH while minimizing the number of parameters. The SFH is defined as a tuple (M$_*$,SFR,$\{t_X\}$) where M$_*$ and SFR are the stellar mass and the star formation rate at the time of observation, respectively. $\{t_X\}$ is an array of lookback times at which the galaxy formed $X$\% of its observed stellar mass.
The number of $t_X$ parameters is best decided using a method such as the Bayesian Information Criterion (BIC) that balances the information available from the observed data with the number of input parameters to prevent overfitting \citep{Iyer2019}.  However, for the ANGST galaxy SEDs, our assumption of across-the-board 10\% photometric uncertainties rendered the BIC suspect due to smaller-than-expected values of $\chi^2$. We therefore utilized a fixed set of five $t_X$ parameters; this remains "non-parametric" in the sense that no particular shape is assumed for the SFHs.     
A galaxy whose SED is being fit with five $t_X$ parameters plus SFR and stellar mass should ideally have more than seven bands to avoid being overfit; the ANGST SEDs have between six and eleven photometric bands per galaxy, with only one galaxy having just six bands, and three with seven bands.  Our models do not come close to spanning the space of all possible SEDs and therefore have significantly fewer than 7 effective degrees of freedom; this is revealed by the best-fit $\chi^2$ from our galaxies remaining above zero even for SEDs with 6 or 7 bands, and that allowed us to  proceed.   

With five $t_X$ parameters, our SFH tuple is (M$_*$,SFR,\{ $t_{16.7}$, $t_{33.3}$, $t_{50}$, $t_{66.6}$, $t_{83.3}$\}). The \{$t_X$\} parameters have a Dirichlet prior indexed by $\alpha$ that allows the user to adjust how correlated the individual $t_X$ parameters are \citep{Leja2017, Iyer2019, Lower2020}.   In \cite{Iyer2019} the optimal $\alpha$ value was found to be 5, with lower values being less correlated and leading to more stochasticity and higher values resulting in a smoother, flatter SFH. We adopt $\alpha=5$ for this work. 
To enable a better reconstruction of the strong early star formation seen in nearby dwarf galaxies \citep[e.g.][]{weisz2, Weisz2014reion}, we relaxed the default Dense Basis prior to allow the SFR to be non-zero at the time of the Big Bang.  
While formally unphysical, this leads to SEDs indistinguishable from those produced by a rapid increase in SFR over the first $\sim$100 Myr after the Big Bang.

The prior on stellar mass set within Dense Basis was changed to have a lower bound of $10^5$M$_\odot$ to make sure that we fully covered the parameter space for our dwarf galaxies. 
For redshift, we placed a narrow prior around the value corresponding to a distance of 10 pc.    
Other factors for each galaxy encoding assumptions for metallicity, dust attenuation, initial mass function (IMF), nebular emission, dust emission, and stellar isochrones, amongst others, are set through FSPS. These are set to the same values as were set for generating the mock photometry as stated in Section ~\ref{section:phot_valid} 
and Appendix~\ref{appendix:systematics}.   
 For a detailed discussion of the handling of these assumptions, see \citet{Iyer2019}.

Dense Basis creates a large atlas of model SEDs generated via random draws from these priors. Each model SED corresponds to an SFH tuple of (M$_*$, SFR, \{$t_X$\}). For our analysis, we evaluate the likelihood of each SED in the atlas with respect to the observed SEDs, and then use them to reconstruct posterior distributions of key parameters, similar to  \citet{Iyer2019} and \citet{Pacifici2012,Pacifici2016a}.  
For each galaxy, these posterior distributions determine a median SFH   along with a 68\% range around the median. In Figure \ref{fig:gp_vs_cmd} we show an example galaxy from our sample with the Mock and Observed SEDs displayed next to the SFHs determined from the CMD and Dense Basis methods.

\section{Results for Individual Galaxies}\label{sec:individual}

Although the CMD and SED methods sample complementary photometric information with different prior assumptions,  there should be common features in the inferred SFHs. \cite{weisz,weisz2} reported diversity in the CMD SFHs for ANGST galaxies.  Taken as a whole, the Dense Basis SFHs agree; they reveal significant variation in the number and time separations of peaks and troughs in the SFR as well as a wide range of specific SFRs. Side by side SEDs and SFHs for individual galaxies are shown in Appendix ~\ref{appendix:individual_sfhs}, ordered by sSFR; while many features in individual SFHs agree between the two methods, there are also  clear disagreements. In this section, we introduce metrics that summarize how well the outputs from these methods compare overall. 

\subsection{Integrated Stellar Mass}
\begin{figure}[h]
    \centering
    \includegraphics[width=0.45\textwidth]{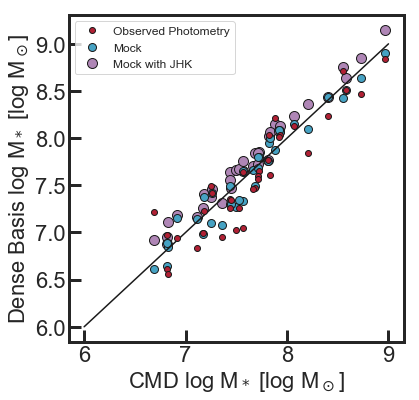}
    \caption{Total stellar mass recovered for the 36 ANGST galaxies by Dense Basis SED fitting for each set of photometry compared with that inferred from the CMD SFHs. The Dense Basis results from observed photometry (red) show the largest scatter and are modestly lower than total stellar masses inferred by CMD SFHs (by 0.08 dex as noted in Table ~\ref{tb:bias}).  Dense Basis results using mock photometry generated directly from the CMD SFH (blue) have less scatter and are nearly unbiased; DB results from mock photometry that includes the \textit{JHK} bands (purple) show the least scatter but overestimate the total stellar mass (by 0.14 dex). 
    }  
    \label{fig:mass}
\end{figure}

We calculate the total stellar mass formed by integrating over the entire star formation history. This is different from the observed stellar mass as the amount of mass retained at the time of observation is approximately 70\% depending on IMF \citep{Conroy2009, conroygunn}. For simplicity, when referring to stellar mass, we mean the integrated star formation since the Big Bang. Figure \ref{fig:mass} compares the total stellar mass formed in the CMD SFHs and the Dense Basis SFHs. In both methods, this property is the most robust to systematics that infer older or younger individual stars due to differences in stellar evolution models or  the handling of degeneracies, although misinterpreting the ages can still modify the inferred stellar mass somewhat since stars of different ages have different mass-to-light ratios. Therefore, we expect good agreement between the methods for this key observable. 

Table~\ref{tb:bias} reports the offset and scatter in this quantity for the observed and mock photometry; we see good overall agreement. Dense Basis results from mocks that include the near-infrared  \textit{JHK} bands show the least scatter but with a 0.14 dex bias. 
Dense Basis results using mock photometry in the available observational bands show almost no bias (0.02 dex underestimate) and moderate scatter (0.14 dex), implying little methodological error in our SED-fitting estimates of this quantity.   
 This allows us to interpret the differences seen between total stellar masses inferred from the Dense Basis results using observed photometry and the CMD SFHs,  where Dense Basis values are offset lower by 0.08 dex and the scatter is 0.22 dex, as resulting from  0.06 dex of systematic disagreement between the two types of observational data with roughly equal contributions to the scatter coming from such disagreements and errors in the Dense Basis estimation.   
To simplify the rest of the figures, we will show only results for the Mock and the Observed photometry.

\begin{table}[h] 
\begin{center}

\begin{tabular}{lrr}
\toprule
Photometry Type &      Offset [dex] &   Scatter [dex]\\
\midrule
Observed                 & -0.08 &  0.22 \\
Mock                     & -0.02 &  0.14 \\
Mock with \textit{JHK}    &  0.14 &  0.08 \\
\bottomrule
\end{tabular}


\end{center}
\caption{Offset and scatter in total stellar mass formed over the SFH reconstructed by the Dense Basis method versus that found by the CMD method, for observed and various mock SEDs.  For mocks, the offset represents a bias in the Dense Basis results; for the Observed SEDs, the offset represents a combination of this methodological bias and systematic differences between the observational data input to the two methods.}
\label{tb:bias}
\end{table}

\subsection{Star Formation Rate}
\label{sec:sfr100}
\begin{figure}[h!]  
    \centering
    \includegraphics[width=0.45\textwidth]{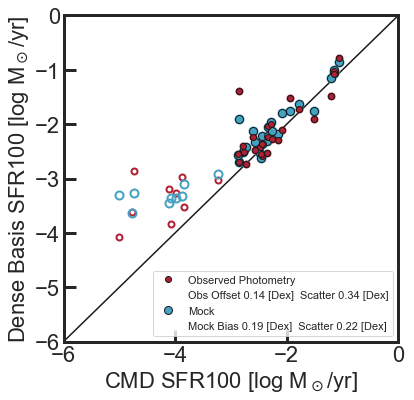}
    \caption{A comparison of the mean star formation rate over the past 100 Myrs for each galaxy. Open circles are galaxies where the uncertainties for the SFR100 estimate from Dense Basis were greater than 2 dex due to very low sSFR, and were dropped from our calculations of offset and scatter for this reason. Dense Basis results from both observed and mock photometry are largely in agreement with those from CMDs, with the exception of the galaxies with very large uncertainties.  As noted in the figure legend, compared to results from mock photomety, results from observed photometry show more scatter but a slightly reduced offset of 0.14 dex. }
  
    \label{fig:sfr100}
\end{figure}

Our SFR calculations are for the mean SFR over the last 100 Myrs rather than instantaneous SFR. This smooths out significant stochasticity in the most recent time bins in the CMD SFHs to match the timescale to which SED-based SFRs are most sensitive. Figure ~\ref{fig:sfr100} shows that 
the SFRs correlate well 
but deviate from a one to one correlation below $10^{-3}$M$_\odot$/yr.

In particular, the galaxies ddo44, ddo71, hs117, and eso294-010 show a much lower SFR in the CMD method than in Dense Basis. This highlights a fundamental difference between what can be inferred from SED and CMD observations. Since the broadband and mock photometry both have SFHs generated from Dense Basis, their combined divergence from the SFR100 of the CMD SFHs highlights differences between the SED and CMD SFH reconstruction methods. For galaxies with extremely low sSFR the signal in the SED is too faint to identify the correct SFR, whereas the CMD method is limited only by the density of the youngest stellar population resolved in the observation. In the case of our SEDs, we see Dense Basis defaulting to the median of the prior and returning large uncertainties for galaxies with such low sSFR that even the rest-UV light is dominated by old stars, making confident SFR measurements impossible with UV-through-NIR SEDs. We show  galaxies with Dense Basis uncertainties on SFR100 greater than 2 dex as open circles in Figure ~\ref{fig:sfr100}.

\subsection{Mass Formation Times}

\begin{figure*} 
    \centering
    \includegraphics[width=0.9\textwidth]{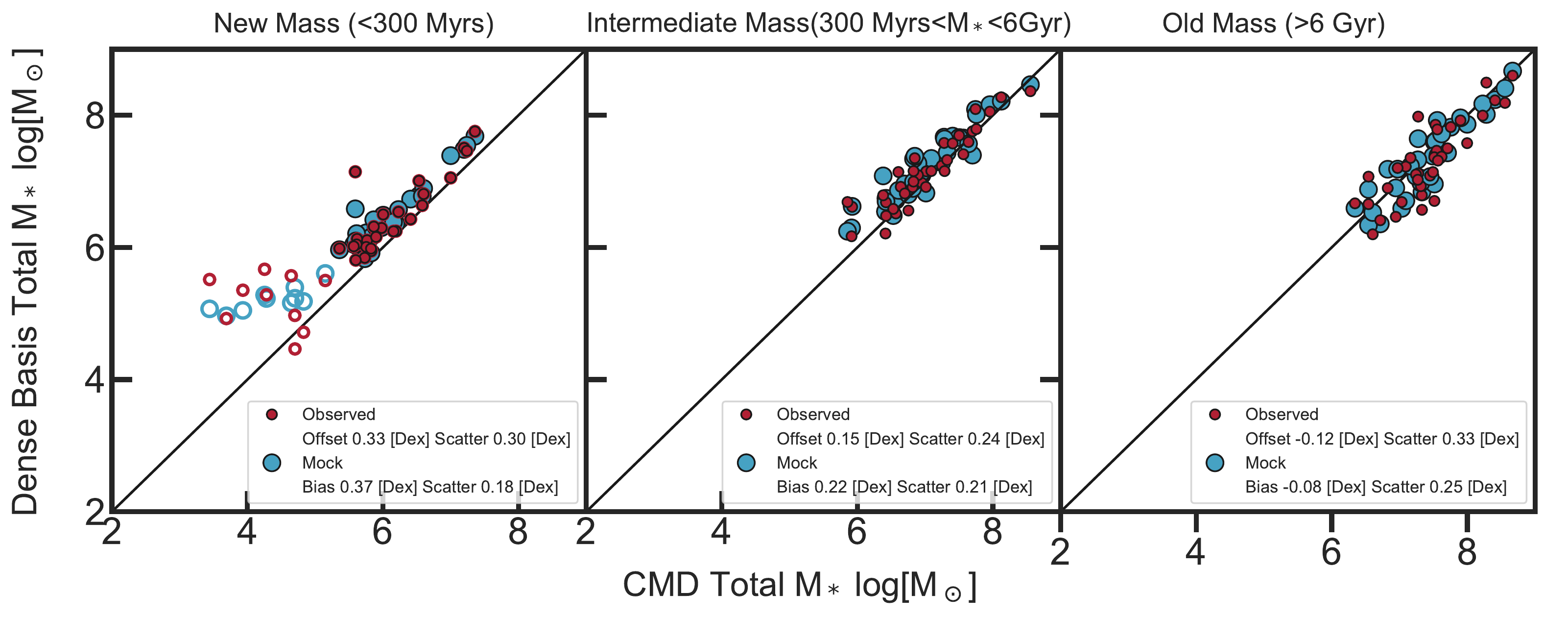}
    \caption{Stellar mass divided into bins of new, intermediate and old mass. In our recent stellar mass bin we find agreement with the exception of the same galaxies with high ($2$ dex) uncertainties in recent SFR100 (indicated with open circles), which we remove from our calculations of offset and scatter in new mass. The discrepancy in new mass for those galaxies is consistent with that seen for SFR100. Intermediate and old mass are in good agreement, with old mass estimates showing more scatter than intermediate. 
    In all three panels, Dense Basis results from observed photometry show modestly increased scatter, likely due to  observational noise, but are otherwise similar to those from mock photometry, implying a lack of systematic differences between information present in the CMDs and observed SEDs.  
    }
    \label{fig:Stellar_mass_3pan}
\end{figure*}

We divided both sets of SFHs into three time intervals, which we define as young, intermediate, and old stellar mass. We choose a definition of young stars based on the logarithmic values of the time bins as those formed in the past 300 Myr. Old stellar mass is defined as having formed between the Big Bang and a lookback time of 6.3 Gyr, as this is the boundary of the oldest CMD SFH bin, and intermediate stellar mass as having formed  in between. In Figure~\ref{fig:Stellar_mass_3pan} we show the results of this analysis. In our young mass panel, we see good agreement with the exception of the same outliers seen in Fig.~\ref{fig:sfr100}. Close inspection of the recent SFHs of these galaxies shows them undergoing a period of quiescence, but with non-zero star formation in the CMD SFHs within the last 300 Myrs. This implies that the CMDs are more sensitive to star formation on such timescales.   With broadband photometry alone, small bursts of star formation are degenerate in the NUV with larger but older bursts \citep{Broussard2019,FloresVelzquez2020}. As noted above, it is difficult to identify small amounts of recent star formation for galaxies with particularly low sSFR, resulting in large SFR uncertainties from SED fitting. 

The intermediate and older stellar mass shows good agreement for the Dense Basis SFHs from both the mock and observed photometry.

\begin{figure*}
    \centering
    \includegraphics[width=\textwidth]{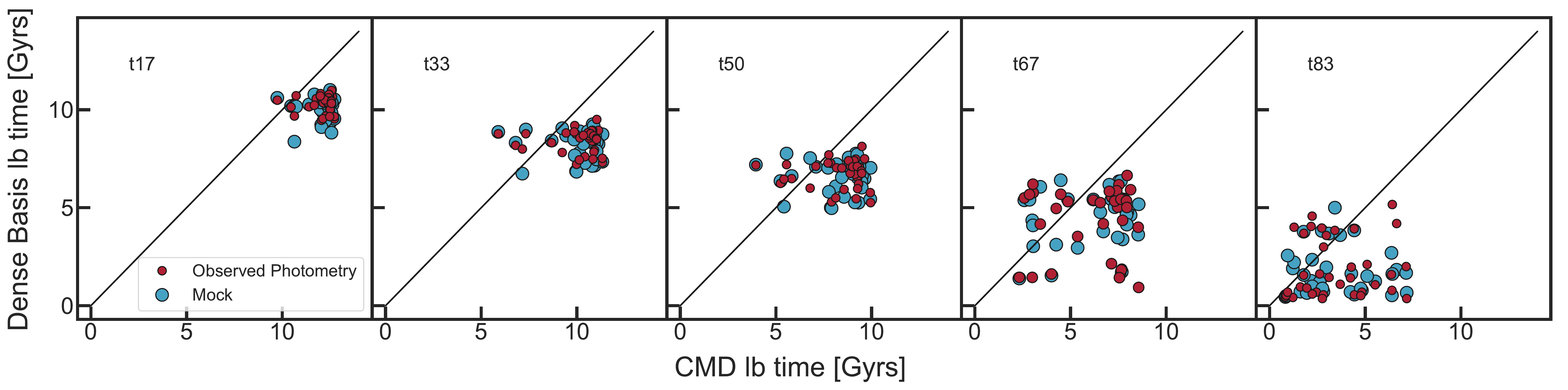}
    \caption{Time of formation of 17\%, 33\%, 50\%, 67\%, and 83\%  of total stellar mass. While the CMDs systematically form mass earlier, there is rough agreement for all mass quantiles, with increased scatter at later times. Values for the offset (bias) and scatter are shown in Table~\ref{tb:tvals bias}.}
    \label{fig:five_panel}
\end{figure*}

 Figure \ref{fig:five_panel} shows the $t_X$ values for both methods in lookback time i.e.,  larger $t_X$ implies an earlier time of formation for that fraction of the mass. We see rough agreement between the methods.  In detail, the CMD method systematically finds earlier mass formation times for most galaxies at all $t_X$, but is most tightly correlated for the earliest $t_X$ with increasing scatter seen with later $t_X$ as seen in Table ~\ref{tb:tvals bias}. This trend is similar to the behavior seen in the comparison of recent mass formed and SFR100. This time the effect is no longer seen with only outliers, but rather suggests there is a systematic where the CMD tends to form mass earlier than the SED method. This systematic may result partially from the   different priors on the shape of the SFHs; in galaxies entering quiescence, the discontinuous prior on the CMD SFH allows for an instantaneous drop in SFR, while the smooth Dense Basis SFHs cannot drop as precipitously.  This is a case where our approach of assuming the CMD SFHs when producing mock photometry and using it to check for methodological bias in the Dense Basis results could prove misleading; if galaxies do not really have SFRs dropping so quickly, the Dense Basis bias might actually be an advantage.  However, since most of the stellar mass is formed at large lookback times, we do not expect that different interpretations of recent quiescent phases can be the primary cause of the overall discrepancy.

\begin{table}[h] 
\begin{center}

\begin{tabular}{lcccc}

\toprule
t$_x$ value & Mock  &{} & Observed &{}   \\
(Gyr) &      bias      &       scatter        &  offset              &    scatter              \\
\midrule
t17    &  -1.99 &      0.87 &      -1.83 &          0.80  \\
t33    &  -1.87 &      1.51 &      -1.71 &          1.45  \\
t50    &  -1.74 &      1.69 &      -1.60 &          1.64  \\
t67    &  -1.61 &      2.10 &      -1.86 &          2.61  \\
t83    &  -1.77 &      2.26 &      -1.59 &          2.29  \\
\bottomrule
\end{tabular}

\end{center}

\caption{Offset (bias) and scatter of fractional mass formation times for Dense Basis output SFHs using mock and observed photometry compared to the t$_x$ values from the CMD SFHs.  The similarity in results between the mock and observed columns implies that we are seeing a methodological bias in the Dense Basis method rather than systematic differences in the CMD and SED data.}
\label{tb:tvals bias}
\end{table}

\subsection{Bursting, Quiescence, and Gaps in Star Formation} 
We will now compare the  fraction of time each galaxy spends in different phases of star formation. Our goal is to define metrics that serve to compare excursions from the recent median SFR for both SFH reconstruction methods in order to provide constraints on  the overall level of stochasticity in individual galaxies and  insight into the evolution of dwarfs. We note that metrics that identify bursting, quiescence, and gaps in star formation are all sensitive to the time resolution of the input SFH, leading to the caveat that metrics such as these would need careful testing for systematics before being used to draw strong conclusions. 

Star formation bursts and periods of quiescence generically refer to a deviation from the mean SFR above or below some threshold; frequently that threshold is a factor of two to three versus the mean \citep[e.g.][]{Hunter1986, Kennicutt2005}. Parameterizing bursts and quiescence is non-trivial, as star formation evolves with cosmic time. For example, \citet{Pacifici2016a} defined the threshold between star forming and quiescent galaxies in terms of specific SFR and the age of the universe. \citet{McQuinn2010} adopted a birthrate parameter modeled on that of \citet{Scalo1986} that they define as two times greater than the average SFR for the last 6 Gyrs. \citet{FloresVelzquez2020} takes a similar approach by setting a threshold above boxcar averages of SFRs of varying short timescales. With these in mind, we define a galaxy to be bursting if its SFR is three times greater than its recent median SFR. Likewise we define a galaxy to be quiescent if it has SFR three times less than the recent median SFR. To avoid having the recent median influenced strongly by ongoing bursts or quiescence, and noting that starbursts in dwarf galaxies can last of order 500 Myr \citep{McQuinn2010},  we add in a delay of 500 Myr. This means that to be considered a burst or quiescent at a given time, a galaxy has to have $\pm$ 3 times its median SFR over the period between 0.5 and 1.5 Gyrs earlier.

\begin{figure*} 
    \centering
    \includegraphics[width=0.7\textwidth]{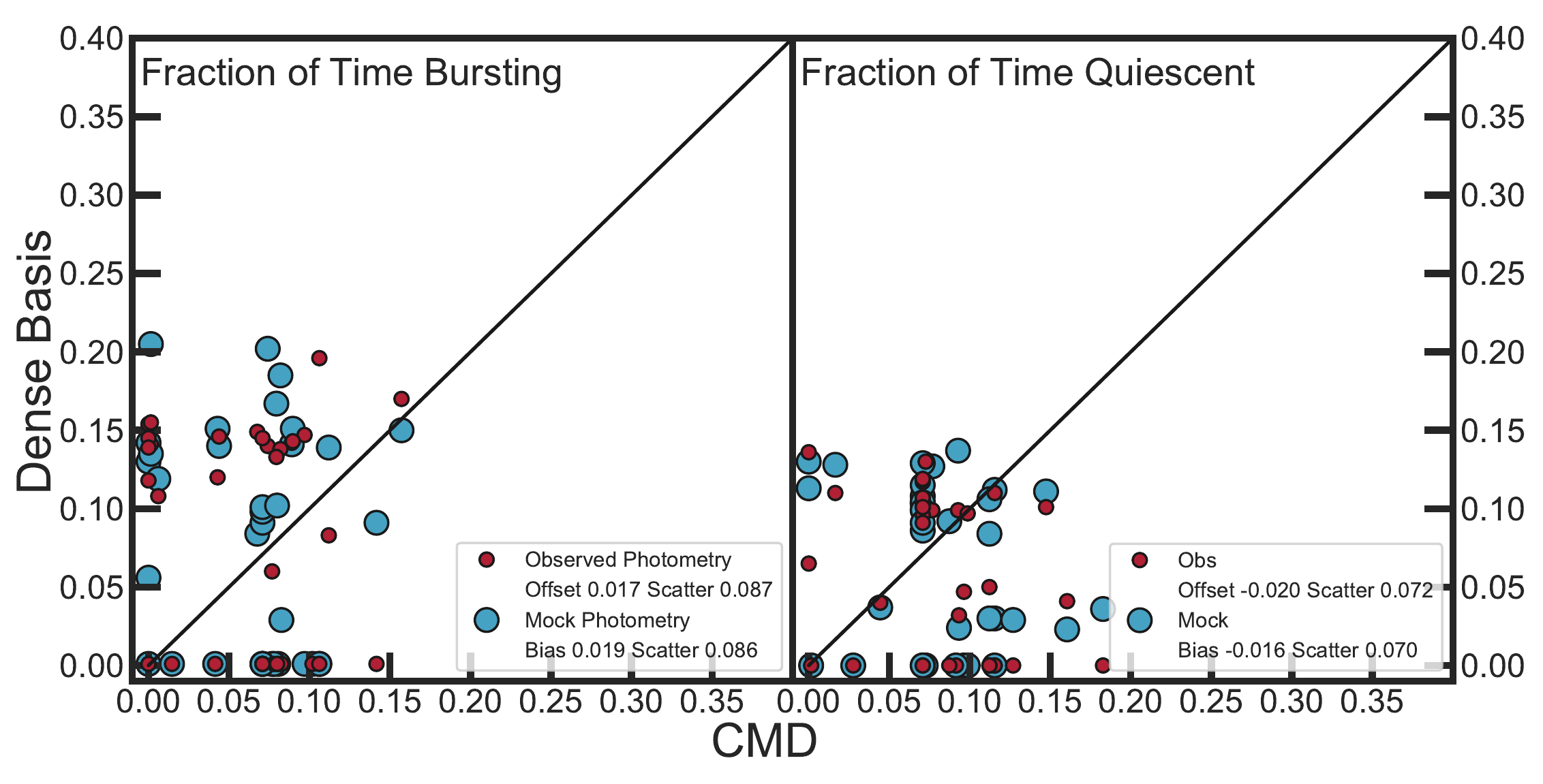}
    \caption{Fractional time each ANGST galaxy spent ``bursting" or  ``quiescent," defined as having SFR three times greater than or less than the median SFR over the period 0.5-1.5 Gyr earlier, respectively. Despite significant scatter, both methods find that galaxies spend $<$20\% of the time either bursting or quiescent.  }
 
    \label{fig:burst_quench_3pan}
\end{figure*}

We show results for this in Figure  \ref{fig:burst_quench_3pan}.  The methods show considerable scatter, but no galaxy spends significantly more than 20\% of its lifetime either bursting or quiescent. For the Dense Basis SFHs, a subset of galaxies never meets this definition for bursting or quiescence, and this is true for a few galaxies' CMD SFHs as well. We note that this analysis takes the binned shapes of CMD SFHs at face value, and that some of the scatter may be caused by the non-differentiable CMD SFHs being compared to the smooth Dense Basis SFHs.   

\begin{figure*}
    \centering
    \includegraphics[width=0.9\textwidth]{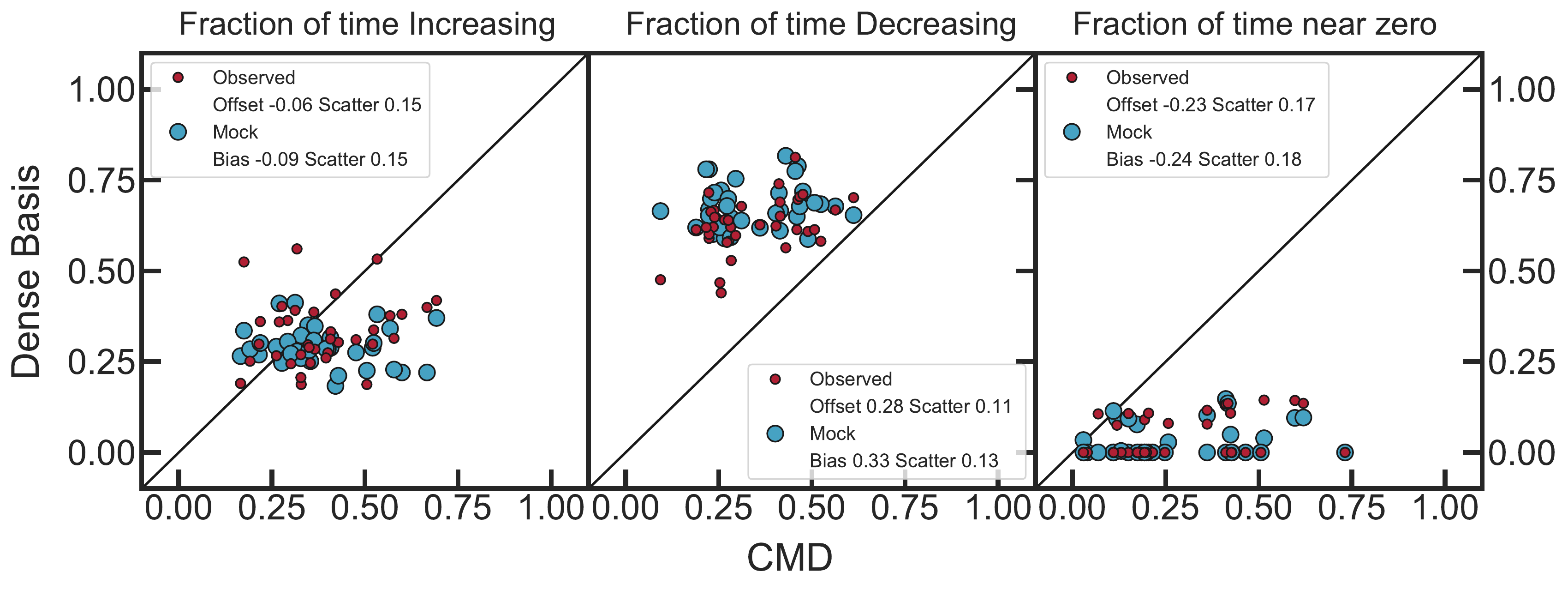}
    \caption{In spite of a large degree of scatter, we find rough agreement in the amount of time the CMD SFHs and Dense Basis SFHs spend at increasing SFR. Results from mock photometry show that Dense Basis tends to overestimate the time spent decreasing and underestimate time spent at SFR$=$0.   }
    \label{fig:time_sf_inc_dec_3pan}
\end{figure*}

It is also worth examining the fraction of time that the 
SFR is climbing or decreasing in an absolute sense. 
The smooth curves of the Dense Basis SFHs are almost always either increasing or decreasing their SFR, although the method allows them to spend brief periods plateaued at SFR$=$0. Quantifying this value for the CMD SFHs is complicated, as they are always at constant slope but discontinuous at bin edges, so for this analysis, we reformulated the CMD SFHs by assuming a constant slope from the center of one time bin to the center of neighboring bins while preserving the stellar mass formed in each bin.  
We also calculated the fraction of time spent with SFR$=$0 for both methods.  
The results in Figure~\ref{fig:time_sf_inc_dec_3pan} show good agreement with a fair amount of scatter for the fraction of time the SFH is increasing. Compared to the CMD SFHs, Dense Basis SFHs systematically overestimate the fraction of time decreasing but spend less time with zero SFR. This appears to be a methodological difference, since the offsets are similar for Mock and Observed Photometry. The discrepancy is likely caused by the different nature of the SFHs; anytime two consecutive CMD SFH bins have SFR$=$0, our diagonal interpolation will yield significant time spent at SFR$=$0, whereas Dense Basis SFHs must decrease smoothly towards zero and, as a result, end up spending little at zero.

\begin{figure} 
    \centering
    \includegraphics[width=0.45\textwidth]{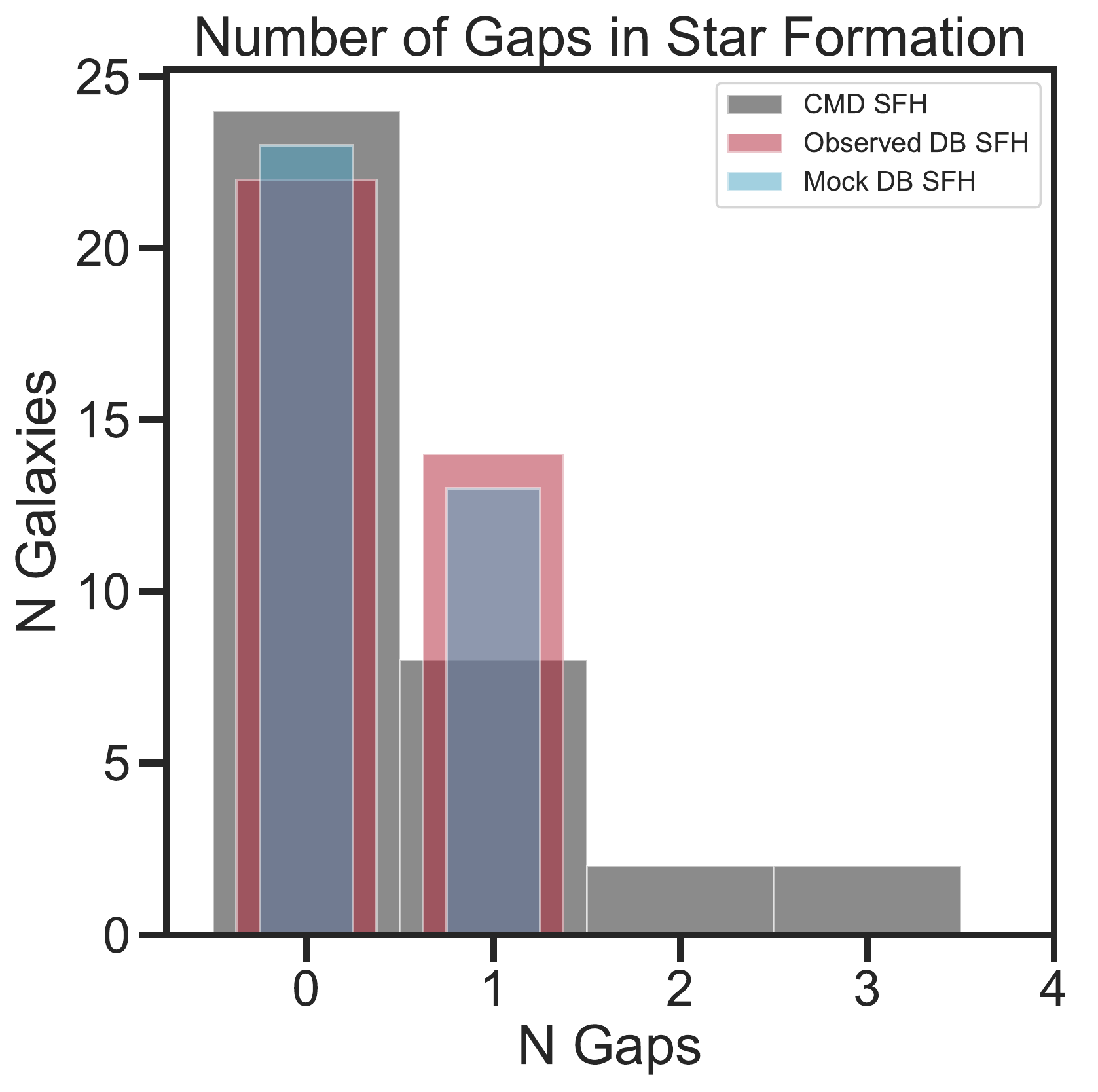}
    \caption{Number of ``gaps'' where the SFR goes to zero in an individual galaxy's SFH. The CMD based SFHs see more gaps for a small number of galaxies. Overall, the histograms agree well.}
    \label{fig:gaps}
\end{figure}

Another metric we used to quantify the overall stochasticity of the CMD and Dense Basis SFHs is to look at the number of periods without active star formation before rejuvenation. These periods, or `gaps', have been noted and quantified previously, in the case of simulated dwarf galaxy SFHs \citep{Wright2018}, but also in observations of dwarfs \citep{Cole2007, Clementini2012, Cannon2018}. These gaps have been hypothesized to be evidence of possible mergers or interactions \citep{Cannon2018,Wright2018, Beale2020}, filamentary gas streams in the IGM \citep{Wright2018, Beale2020}, or a long cooling timescale for gas in a warm, diffuse halo \citep{Cole2007}. Here we determined the number of gaps by counting every time the SFH went from having a non-zero SFR to having zero SFR. Our expectation was that the CMD SFHs would display more gaps than Dense Basis by this metric, since the binned, discontinuous nature of these SFHs allows for zero SFR to be assigned to any bin.   However, for the ANGST dwarf galaxies as a population, the methods were found to be in generally good agreement, as seen in Figure~\ref{fig:gaps}. Nonetheless, analysis of individual galaxies showed that the two methods found the same number of gaps in only about 50\% of the individual galaxies.

\section{Comparison of Median Normalized SFHs}
\label{sec:median}

With the initial goal of finding systematic features in the Dense Basis and CMD SFHs that might reflect biases in either method, we normalized each output SFH and then took a median over all ANGST galaxy SFHs at each point in time.  These ``median normalized'' SFHs are similar to the ``scaled median'' stacks of photometry recommended for SED-fitting by \citet{Vargas2014}. The motivation is similar; in an overall median, the most (least) massive galaxies tend to have high (low) SFRs at each point in time, leaving the median SFH to be dominated by a small set of medium-mass galaxies.  Normalizing the SFHs before taking the median allows us to determine the typical SFH {\it shape}.   

\begin{figure*} 
    \centering
    \includegraphics[width= 0.75\textwidth]{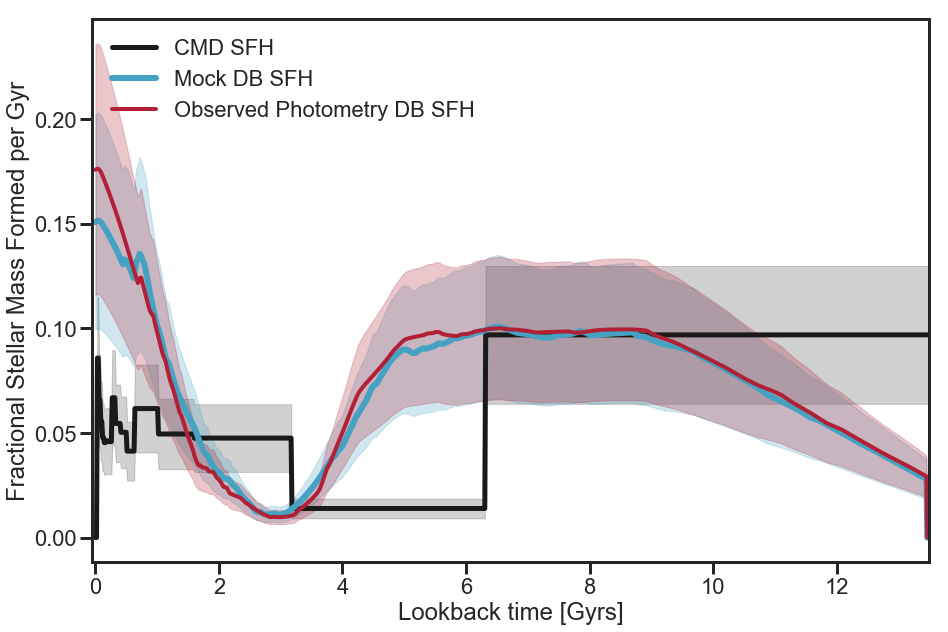}
    \caption{Median Normalized SFHs from the CMD and Dense Basis methods for both the observed photometry (red) and the mock photometry (blue) in units of the fraction of the eventual total stellar mass formed per Gyr. Shaded regions indicate 68\% scatter around the median. The surprising agreement of a descent into quiescence at 4-6 Gyrs lookback time followed by a rejuvenation of star formation in the last 2-3 Gyrs provides evidence of conformity in the SFHs of Local Volume dwarf galaxies.}
    \label{fig:med_sfh_all}
\end{figure*}

In Figure \ref{fig:med_sfh_all} we show the  median normalized CMD and Dense Basis SFHs.   
The median normalized CMD SFH shows a period of quiescence starting 6 Gyrs ago ending with new star formation that began 3 Gyrs ago and has continued to the present day. The Dense Basis SFHs 
from observed photometry show the same events but with quiescence and rejuvenation  both starting $\sim$1 Gyr later and recent SFRs $\sim$3 times higher.   
 The median normalized Dense Basis SFHs from mock and observed photometry show almost perfect  agreement, implying that the differences in timing and recent SFR versus the median normalized CMD SFH are due to biases in the Dense Basis method.  This means that all results are consistent with the timing of quiescence and rejuvenation events found in the CMD SFHs.  The same features are present in median SFHs that have not been normalized, but normalizing each SFH before taking the median provides a more robust measure of overall shape. Additionally, dropping the galaxies with high uncertainties in SFR fails to affect this result with the exception of increasing the recent SFR by a factor of $\sim1.5$.
Because SFHs reconstructed from CMDs and SED fitting are sensitive to different observational systematics, finding agreement between them at this level provides strong evidence that the observed trends are robust.  

In Appendix~\ref{appendix:systematics} we illustrate the degree to which changing the assumptions for dust, metallicity, isochrones, stellar 
models, and TP-AGB treatment influence the results shown in Figure~\ref{fig:med_sfh_all}.  While modest quantitative changes appear, the median normalized SFHs produced by SED-fitting are impressively robust to these potential systematics.

Despite the reported diversity of individual galaxy SFHs, a visual inspection of individual SFHs as in Appendix ~\ref{appendix:individual_sfhs} reveals that, within uncertainties, 25 out of the 36 ANGST galaxies show features matching those revealed in the median normalized SFHs. These features suggest a correlation in star formation at different times between galaxies separated by up to 8 Mpc, well beyond the 1-halo regime where conformity is expected.

\subsection{Median Normalized SFHs of Galaxy Samples Split by sSFR}
 
\begin{figure*} 
    \centering
    \includegraphics[width=0.95\textwidth]{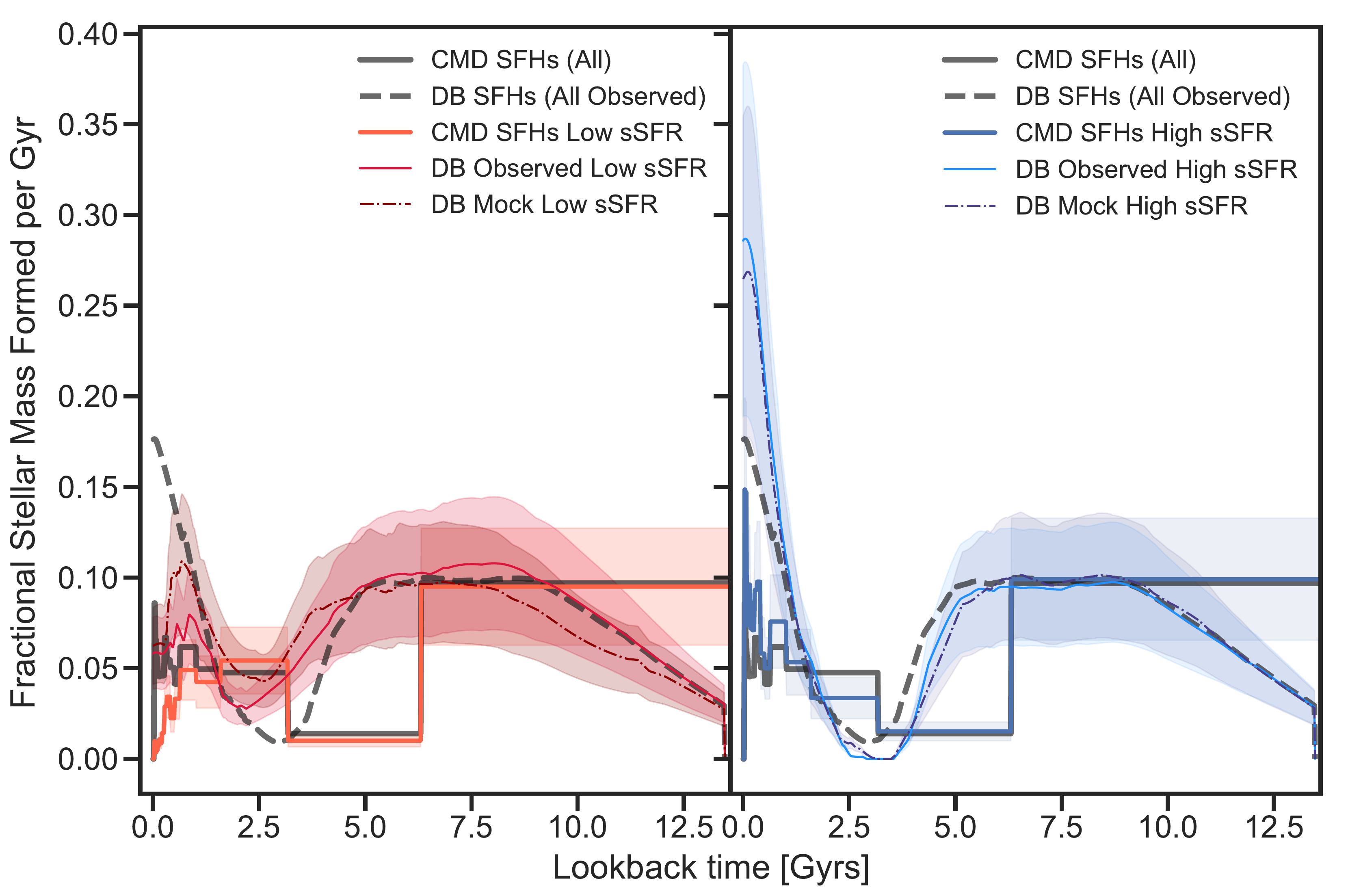}
    \caption{Splitting our sample in two at median SFR100/M$_*$ of $10^{-10}$ yr$^{-1}$, we look at galaxy samples with low versus high sSFR. In the left panel, we see that the median normalized SFHs from both CMDs and Dense Basis are in general agreement with the overall medians but show lower fractional stellar mass formed near the time of observation, as expected given the sSFR selection.  The high sSFR sample in the right panel also shows good agreement with the overall medians except for the expected increase in recent star formation.  
    Both sSFR-selected samples show quiescence and rejuvenation at the same epochs identified in the overall medians.}
    \label{fig:sSFR_all}
\end{figure*}

 The agreement in the median normalized SFHs led us to examine whether or not the median could be dominated by a subset of rejuvenated galaxies. To identify the rejuvenated galaxies, we cut our sample on median SFR100/M$_*$. As current SFR100 is a more robust SFR indicator than instantaneous SFR, we used this to calculate a specific star formation rate (sSFR).  We divided the sample in two at the median sSFR value of $10^{-10}$ and then plotted the median normalized SFHs of these subsamples in Figure \ref{fig:sSFR_all}. 

For the low sSFR sample, Dense Basis remains consistent for the older star formation and matches the median in its episode of quiescence followed by rejuvenated star formation. During the episode of quiescence, the SFR did not drop as steeply as the overall median, and the minimum was higher. The rejuvenated star formation lasted until 1 Gyr ago, but then the SFR declined again. The low sSFR sample of CMD SFHs remains consistent with the overall median until the SFR declined 1 Gyr ago, consistent with the sSFR selection of this sample.

The high sSFR median normalized Dense Basis SFH tightly followed the median of the entire sample. The quiescent episode dipped down to SFR=0 before increasing again to final SFRs higher than the median. Likewise, the high sSFR median normalized CMD SFH agreed with the overall median until 3 Gyrs ago, when the SFR dropped below the overall median. 
Star formation then increased, and this high sSFR CMD SFH rose to the expected higher SFR at the time of observation.

\subsection{Median Normalized SFHs over Regions}

\begin{figure*} 
    \centering
    \includegraphics[width = 1 \textwidth]{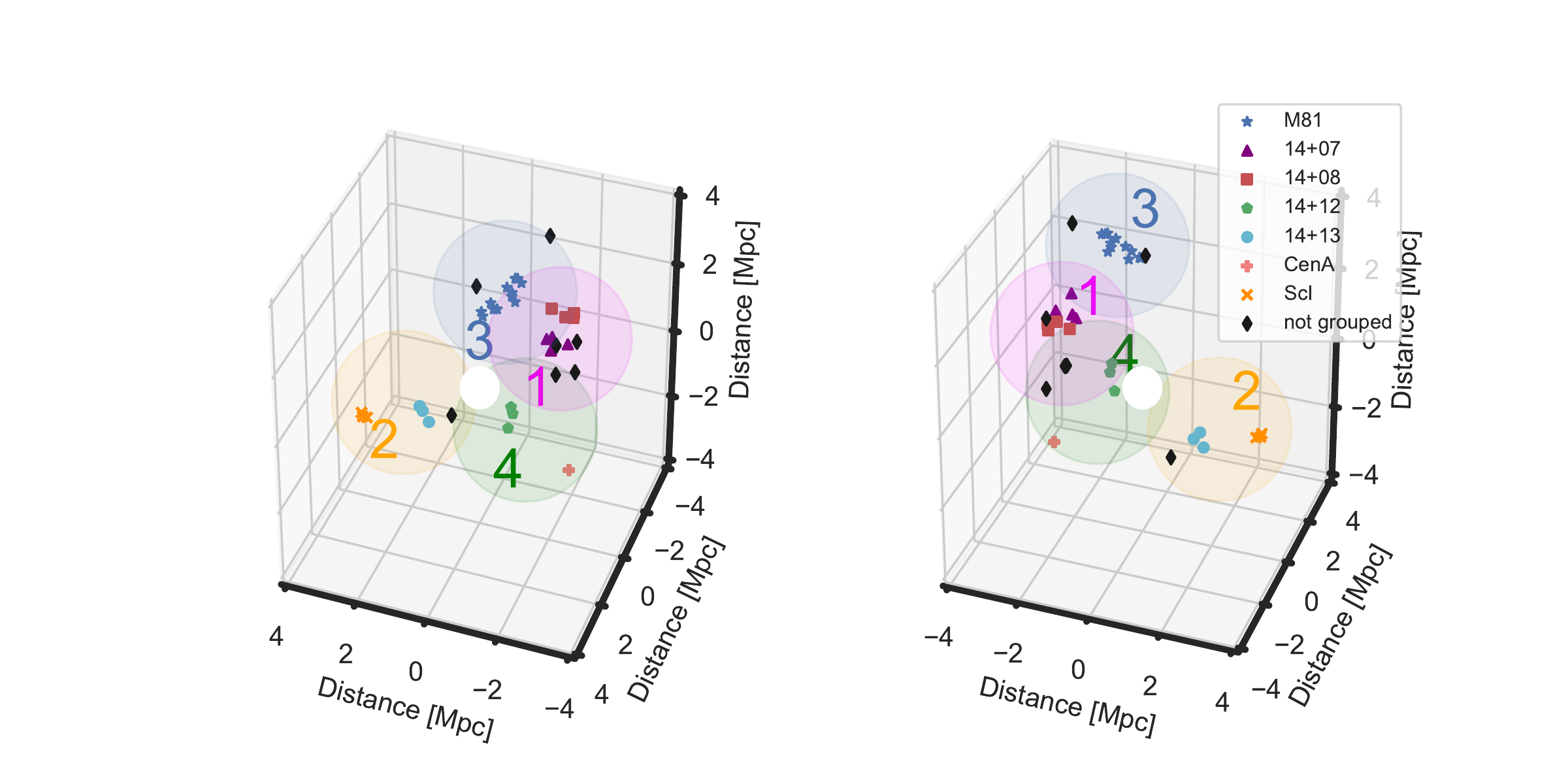}
    \caption{The galaxies in our sample centered around the Milky Way, which is represented by the white sphere at center. Field galaxies are marked as black diamonds, while other shapes and colors indicate known groups. We sort these galaxies spatially into four regions based on a KMeans clustering analysis. The different colored spheres represent these four regions. Legend labels which start with 14+ indicate groups  identified by common right ascension.  
    The left and right panels show two different projections to illustrate that each ANGST galaxy lies in exactly one of the regions.}
    \label{fig:gals_by_region}
\end{figure*}

To check if a particular region of the Local Volume might be driving the shape of the median normalized SFHs, we split the volume containing the 36 ANGST dwarf galaxies to see if different regions within the Local Volume exhibit similar SFH shapes.  The M81 group is prominent in our subsample and consists of 12 galaxies in our sample of 36.  Several other groups contribute 3-4 galaxies each to our sample. Since we are looking at the median behavior of normalized SFHs, it seemed prudent to look at regions more populated than the smaller groups in order to have samples large enough for the median to be robust. Using a KMeans clustering analysis, we defined four main regions where our galaxies lie. Each of these regions may be dominated by members of one or more groups, but may also include field galaxies. A full description of group memberships of the ANGST sample can be found in \citet{dalcanton}.
We show all of the galaxies in our sample in Figure \ref{fig:gals_by_region}, where colored spheres show how we divided the sample into regions.

\begin{figure*} 
    \centering
    \includegraphics[width=0.95\textwidth]{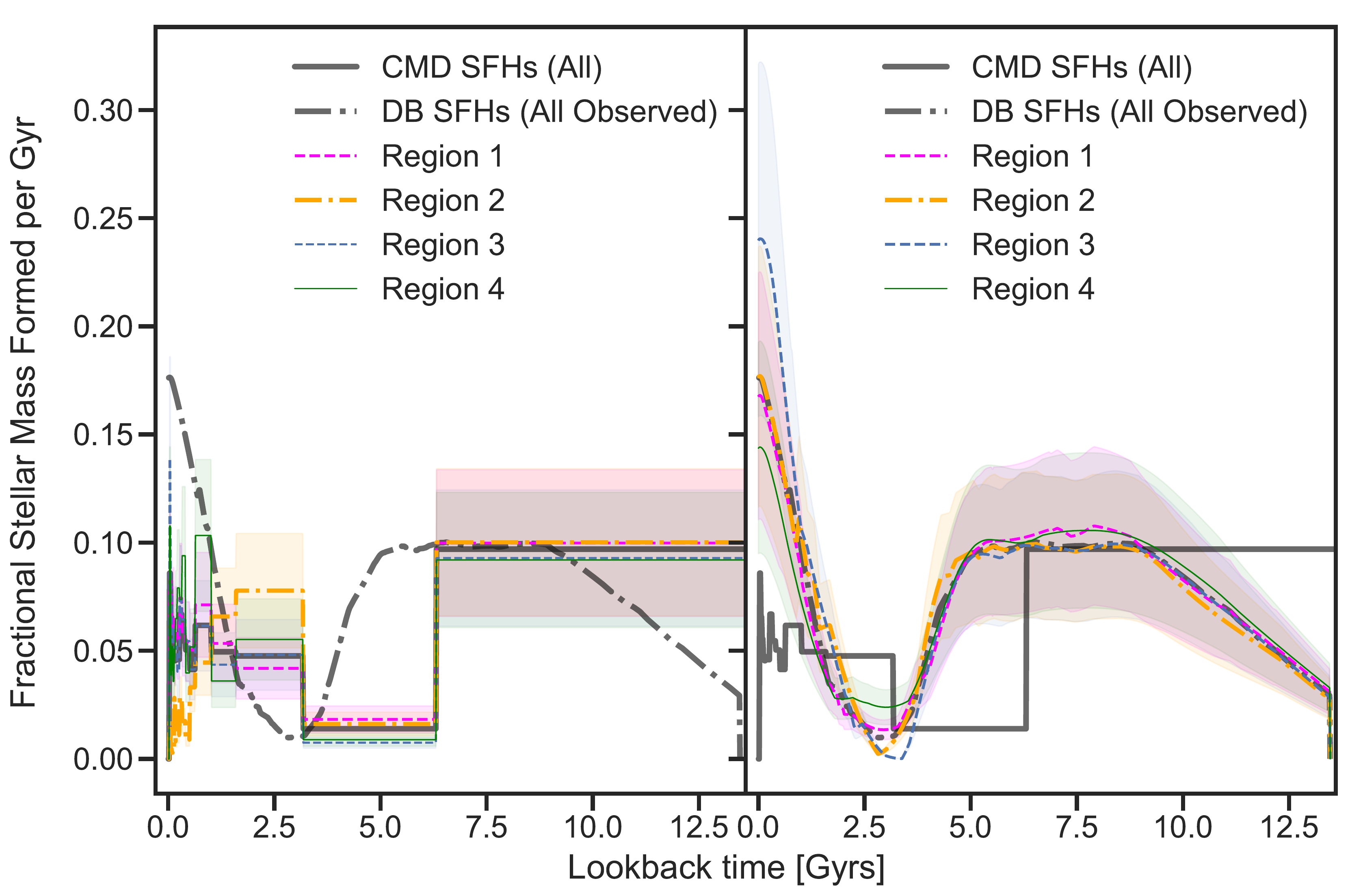}
    \caption{Median normalized SFHs for the regions shown in Figure \ref{fig:gals_by_region}. The left hand panel shows the median normalized CMD SFHs for each region with colors corresponding to those in the previous figure. The gray solid (dashed) line shows the median normalized CMD (Dense Basis) SFH for the full sample. In the right panel, the Dense Basis sample is broken down into regions as labeled.  
    No particular region or group appears to drive the overall behavior of the median normalized SFHs; 
    all four regions show quiescence and rejuvenation at the same epochs as the overall medians.}
    \label{fig:med_reg}
\end{figure*}

If the features seen in the median normalized SFHs are driven by environmental effects from a particular group, we should see that signal in the median normalized SFH for the region containing those group members. In the left-hand side of Figure \ref{fig:med_reg} we show the median CMD SFHs of galaxies broken down into the four regions. The trend towards quiescence  at 6 Gyrs lookback time remains, as does the period of new star formation starting around 3 Gyrs ago, although Region 1 shows reduced star formation in the past Gyr. 
We also create median normalized SFHs for each region using the Dense Basis SFHs from observed photometry, as shown in the right-hand side of Figure \ref{fig:med_reg}, and find similar behavior. There is still good agreement between regions with an episode of quiescence 3-4 Gyrs ago followed by an increase of star formation to the time of observation.

\section{Conclusions}
\label{sec:Disc}

 
Because the CMD and Dense Basis methods of SFH reconstruction impose very different prior assumptions on the shape of SFHs,  direct comparison by eye can be challenging. When one SFH is slowly varying and smooth, while the other is binned and fluctuates significantly, it can be unclear which case more closely matches the ``truth.'' This makes it crucial for methods to be compared against a ground truth model, as in \cite{Rathi2020}, where SPH N-body simulations were used to create mock CMDs, allowing for a side by side comparison of SFHs from the CMD method with the input SFHs. 
Nevertheless, our comparison of these two methods through the metrics introduced in this paper has illuminated robust areas of agreement as well as a few systematics that generate differences between the CMD and SED methods. 

We used the CMD SFHs as inputs to generate mock photometry and analyzed the resulting SEDs with the Dense Basis code.  This provided validation, with the resulting differences in SFHs identifying systematic biases of the Dense Basis method when applied to these low-redshift dwarf galaxies that should also affect the Dense Basis results for their observed photometry.   
Differences between the Dense Basis SFHs inferred from the mock and observed photometry then offer insight into systematic differences between the CMD and SED data.
We introduced a set of metrics for comparing SFHs and found that, in most cases, the Dense Basis results showed similar offsets versus CMD SFHs for the mock and  observed photometry, implying that the CMDs and observed SEDs contain similar information about these galaxies' SFHs.  The Dense Basis results for observed photometry show greater scatter versus the CMD SFHs, likely due to the presence of observational noise.

 Our SED-based recovery of the total stellar mass formed in each galaxy was consistent with the total stellar mass inferred from CMD SFHs. Dense Basis had little bias using mock photometry generated from the CMD SFHs without near-infrared photometry probing the 1.6 $\mu$m bump, with increased bias of 0.14 dex but reduced scatter if the \textit{JHK} bands were included. 
 Using the observed broadband photometry, Dense Basis had an offset of 0.08 dex and scatter of 0.22 dex versus the CMD SFHs. 
 
 The determinations of SFR over the past 100 Myr, SFR100, match well between the two methods for most galaxies. For a set of ANGST galaxies with very low specific SFR, the CMD method finds SFR close to $10^{-5}$ while Dense Basis finds SFRs between $10^{-4}$ and $10^{-3}$.
 For galaxies with especially low sSFR, the Dense Basis code returns SFR100 values close to the median of the prior along with large uncertainties.
 As discussed in our results for SFR100, this is likely due to even the UV light being dominated by older stars at such low sSFR, making a broad-band SED determination of SFR difficult.

 There was good agreement for total stellar mass recovered from the oldest times to 300 Myr ago, but the more recent stellar mass formed shows a systematic difference between the methods.  Dense Basis results for recently formed stellar mass have an offset of 0.33 dex and 0.37 dex for observed and mock photometry respectively, after excluding the set of galaxies with very low specific SFR and correspondingly high uncertainties in recent SFR.



Comparing times at which fractions of the total stellar mass were formed ($t_X$) revealed qualitative agreement between the two methods but with the Dense Basis times biased to be almost 2 Gyr too recent for all fractions.  For Dense Basis results from both mock and observed photometry, the scatter versus the CMD SFHs gradually increased as the fractions increased.  The agreement between the CMD SFH and Dense Basis SFH quantiles at earlier times appears to reflect the agreement between the flat shapes of the SFHs from the Big Bang to 6 Gyrs ago. After this time, the SFHs become more diverse for both methods; the 2 Gyr offsets are caused by a methodological bias in the Dense Basis results towards greater recent star formation in these galaxies.  

Our metrics found that, for both methods, the ANGST galaxies spent $<20$\% of the  time bursting, defined as having SFR at least 3 times greater than the median SFR over the period 0.5 to 1.5 Gyr earlier.  The galaxies also spent $<20$\% of their time in  quiescence, defined as having SFR less than 1/3 of the same earlier median SFR.  Results were generally in good agreement for the overall fraction of time spent with increasing and decreasing SFR. 
When measuring statistics for the sample as a whole, the number of episodes per galaxy where the SFRs are zero i.e., gaps, compare favorably between the two methods. Dense Basis predicts more galaxies with one gap, while the CMD SFHs have some galaxies with two or three gaps. However, the methods only agree on the number of gaps for individual galaxies about half the time.  
For all of these newly introduced metrics for comparing SFHs, we found the agreement good enough to imply that the CMD and SED methods are revealing consistent information.  We caution the reader that a more detailed investigation of the systematic effects of variable time resolution over the course of cosmic time in both the CMD and SED fitting methods would be needed before interpreting disagreements in these metrics as true discrepancies.

To study similarities and differences in SFH shape between the CMD and Dense Basis methods, we compared median normalized SFHs for the 36 ANGST dwarf galaxies.  The methods differ by $\sim1$ Gyr in the exact timing of events and the amplitude of ongoing star formation, but the Dense Basis mock photometry results imply these to be  methodological biases in the Dense Basis results for these galaxies rather than a true disagreement. An investigation of potential SED fitting systematics caused by the choice of metallicity, dust law, isochrones, stellar models, and TP-AGB treatment 
(see Appendix~\ref{appendix:systematics})
found minimal impact on the median normalized Dense Basis SFHs.  The best interpretation of the median normalized SFHs therefore follows the CMD SFH timing, which shows a period of quiescence at 3-6 Gyr lookback time followed 
by rejuvenated star formation over the past 3 Gyr.  The median of all SFHs without normalization revealed the same features but is a less robust measure of SFH shape. 
We split the sample on sSFR, calculated the median normalized SFHs, and found these same features present in both sub-samples. 
We then divided the sample into four regions to check whether the environmental effects of a particular group might be dominating these results, but found that the features agree well for all regions.

The motivation for this work was to find how well the Dense Basis SFH reconstruction technique works at low redshift, and to provide a set of metrics that enable comparison between different methods. While mostly concordant, the metrics used here have illuminated some fundamental disagreements between the two methods that imply methodological biases in the Dense Basis approach when applied to these low-redshift dwarf galaxies. A symmetrical investigation of the systematics for SFHs determined from CMDs is beyond the scope of this paper but was recently undertaken by \citet{Rathi2020}. We also looked for systematic disagreements between the Dense Basis results from the observed photometry and those from the mock photometry produced by assuming the CMD SFHs, but nothing significant was found, implying that the CMDs and SEDs reveal consistent information about the ANGST galaxy histories.  

This compatibility posits that higher S/N measurements of galaxy SFHs could be provided by the development of hybrid methods that combine multi-wavelength broadband photometry with color magnitude diagrams.  Possible 
hybrid methods include the pixel CMD technique outlined by \cite{Cook2020} or simultaneously fitting stellar population synthesis models to combinations of integrated SEDs and fully-resolved CMDs covering matched regions of sky. Hybrid methods can probe the semi-resolved galaxy regime, exploit the fact that SED- and CMD-based SFHs are sensitive to different systematics, and mine information about galaxy SFHs from a mix of fine spatial resolution and a broad wavelength range. 

Further study of this sample with a hybrid method could facilitate expanding robust star formation history reconstruction to a sample of galaxies at distances greater than 4 Mpc where deep CMDs may not be readily available, even with JWST. Local galaxies offer a special opportunity to constrain environmental factors and feedback that are poorly constrained at high redshifts.

Since our sample of 36 dwarf galaxies is diverse in environment and morphology, we did not expect to see evidence of coherent events in their SFHs. However, the normalized median SFHs revealed periods of quiescence and rejuvenation that are uncorrelated with sSFR and were found to be common across 4 separate regions of the (D$<$4 Mpc) Local Volume, with 25 of the galaxies in our sample showing consistent features in their individual SFHs. This is especially unexpected considering the separations of several Mpc. 
Finding conformity at this scale implies the two-halo conformity described in \cite{Kauffmann2012:1209.3306v2} and motivates further observations and simulations of the Local Volume. 
To the best of our knowledge, this conformity has not been predicted by galaxy formation models, including cosmological hydrodynamical simulations.  
For the finding to be spurious, it would require matching systematic flaws to exist in the CMD and SED-fitting methods for SFH reconstruction, despite the fact that they are sensitive to complementary systematics.  

This exciting revelation about the nearby universe merits further investigation of 
 possible ``special epochs" in the history of the Local Volume.  One approach  would be to  study the SFHs of more massive galaxies at slightly larger distances to see if they show the same periods of quiescence 3-6 Gyr ago followed by rejuvenation over the past 3 Gyr 
that we have found for the ANGST galaxies within 4 Mpc.

 \section{Acknowledgements}
 The authors would like to thank Dan Weisz for kindly providing the high resolution CMD SFHs for ANGST galaxies, and Yao-Yuan Mao for their insightful comments. 
 This material is based upon work supported by the U.S. Department of Energy, Office of Science, Office of High Energy Physics Cosmic Frontier Research program under Award Numbers  DE-SC0011636 and DE-SC0010008, which supported CO, EG, and AB. 
 
 \clearpage


\appendix

\section{Discussion of Systematics} 
\label{appendix:systematics}
Systematics play a major role in both SED SFH reconstruction and CMD SFH reconstruction. We have done our best to match the model assumptions used to create the CMD SFHs, 
but we can investigate how changing these assumptions would influence our results. Table \ref{tb:fsps} shows specific FSPS settings that were changed from the Dense Basis defaults to those used in \cite{johnson} to best match the CMD model assumptions.

\begin{figure}[H]
    \centering
    \includegraphics[width=0.75\textwidth]{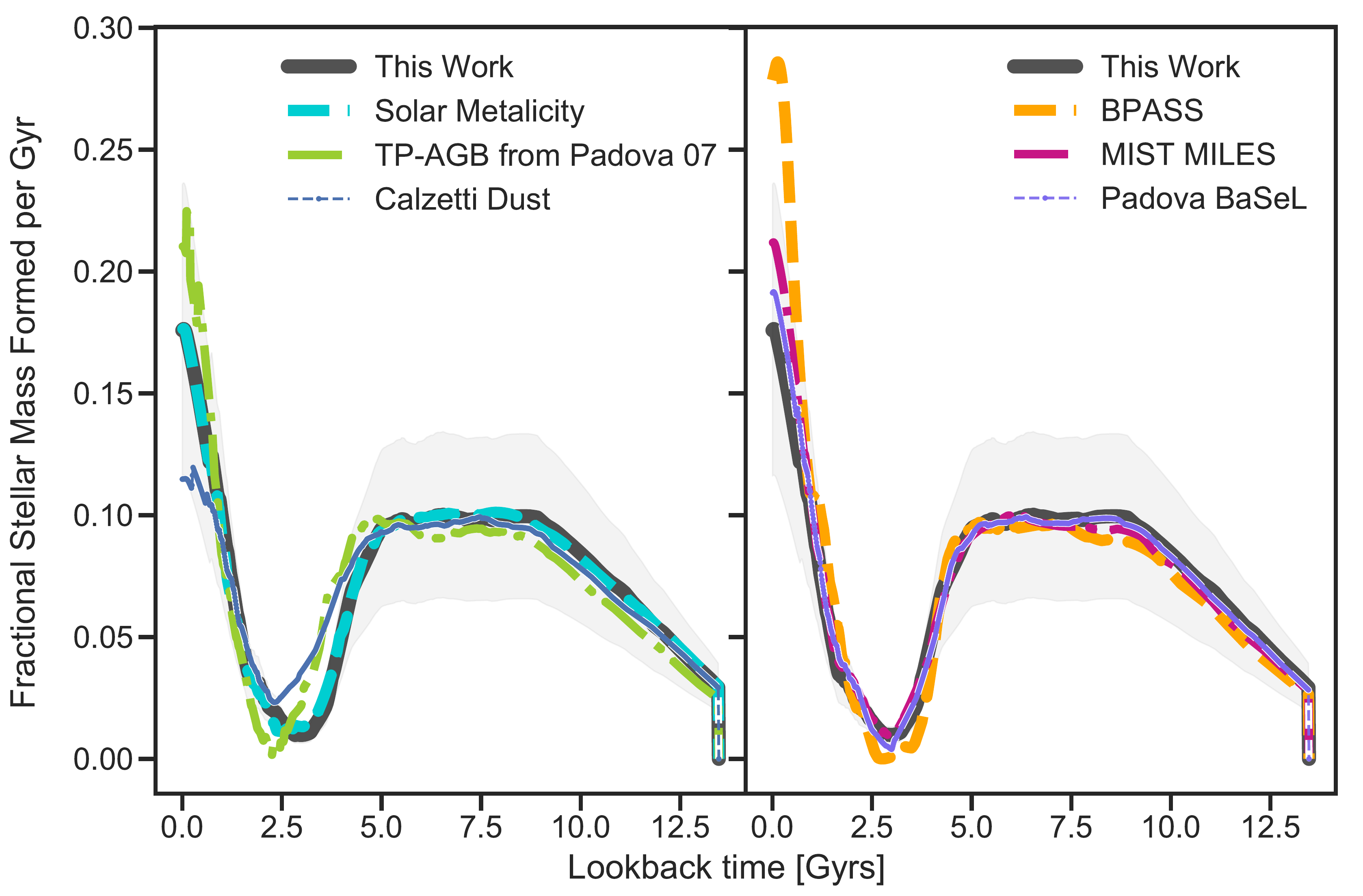}
    \caption{Systematics comparison showing  median normalized SFHs with different model assumptions.  Our model, shown in the black curves in both panels, has custom metallicity values as output from the CMD Method, a TP-AGB normalization of \cite{Villaume2015} as implemented by FSPS, Padova isochrones \citep{Bertelli1994, Girardi2004, Marigo} and MILES \citep{Sanchez-Blazquez2006} spectral library. On the left the cyan curve shows the normed median when all galaxies are assigned solar metallicity, the green shows a Padova 07 TP-AGB normalization, and the dashed blue curve shows results assuming the Calzetti dust law. We see moderate changes from TP-AGB and dust treatments, with Padova 07 predicting a more recent quiescent period and increased recent star formation whereas Calzetti dust produces an equally recent but weaker trough and reduces the recent star formation. On the right we compare with different stellar libraries.  Changes are minimal, with the exception of BPASS leading to an increase in recent star formation. }
    \label{fig:systematics}
\end{figure}

We use the metallicities for each galaxy as calculated by the CMD method for consistency, but as a check we run our analysis with solar metallicity as shown in the left panel of Figure \ref{fig:systematics}. 
FSPS allows for different normalizations to account for thermally pulsating asymptotic giant branch (TP-AGB) stars within the isochrones. For our initial analysis we implemented the most recent normalization from  \cite{Villaume2015}. Since degeneracies caused by TP-AGB stars are seen between 2-3 Gyrs in the SFH, we can change this normalization to see the effect. In the left panel of Figure \ref{fig:systematics} we show our results with the older Padova 2007 \citep{Marigo} normalization. 
The dust law used in our analysis was a custom power law based on \cite{johnson} meant to match the CMD method dust assumptions. Our results with the significantly shallower Calzetti dust law \citep{Calzetti2001} are shown in the left panel of Figure \ref{fig:systematics}.
All of these have minor effects on our results, the least being metallicity. TP-AGB treatment is more significant in moving the timing of the quiescent period and increasing the SFR100. Changing the dust law has a more noticeable effect, as star formation is not as deeply suppressed at 2.5 Gyrs and is lower at the time of observation. 

In the right panel of Figure \ref{fig:systematics} we compare our results with different choices for isochrones and stellar libraries. Our initial results were run with Padova \citep{Bertelli1994, Girardi2004, Marigo} isochrones and MILES \citep{Sanchez-Blazquez2006} spectral libraries. In the right panel we show our results using BPASS \citep{Eldridge2012}, which takes into account binaries in stellar models and isochrones. We also use different combinations of isochrones and stellar libraries i.e.,  MIST \citep{Dotter2016, Choi2016} with MILES and Padova with BaSeL \citep{Lejeune1997}. 
Different combinations of stellar libraries and isochrones made little difference to the normalized median result with the exception of the models that account for binary stars, leading to a modestly higher recent star formation rate.

\begin{table}[H]
\begin{tabular}{c|c|c|c|c|c|c}

    FSPS param & imf\_upper\_limit & imf\_lower\_limit & dust\_type & add\_dust\_emission & add\_neb\_emission & dust1 \\
    \hline 
    \hline
    Dense Basis & 120 & 0.8 & 2 & False & False & 0\\
    Johnson13 & 100 & 0.1 & 0 & True & True & 0.23\\
    \hline
     FSPS param & dust\_index & dust\_tesc & dust1\_index & duste\_gamma & duste\_qpah & \\
    \hline
       \hline 
    Dense Basis & -0.7 & -1 & 7 & 0.01  & 3.5 & \\
    Johnson13 & -1.3 & -1.3 & 7.7 & 0.001  & 0.5 & \\
    \hline
 
\end{tabular}

\vspace{0.1 in}
\caption{Specific differences between default FSPS parameters used by  Dense Basis, and the parameters run in \cite{johnson}. Parameters are easily changed in FSPS.}
\label{tb:fsps}
\end{table} 
\vspace{0.1 in}

\section{Galaxy CMDs with Star Counts and SEDs}
\label{appendix:hst_gst}
\begin{figure}[H]
    \centering
    \includegraphics[width=0.9\textwidth]{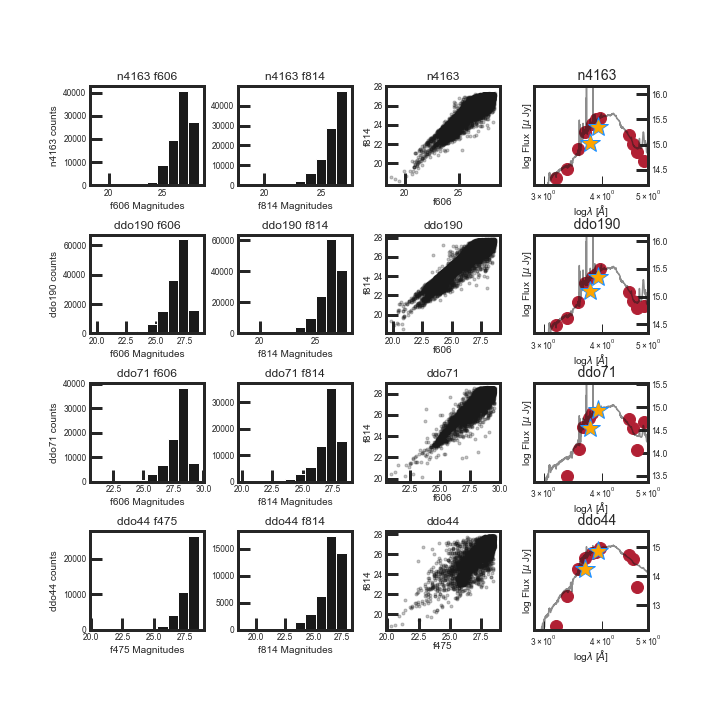}
    \caption{An example of four galaxies for which the observations were not mosaicked and fluxes could be easily summed to create extra bands. In each row, histograms are shown in the first two panels for HST filters, followed by these observations plotted against one another. The last panel shows the broadband photometry in red over the spectra generated from the CMD SFH in black, 
    with photometry from the simple summed filters in orange. There is general agreement, but the offsets are significant enough to disfavor including these additional bands in our analysis. }
    \label{fig:gst_1}
\end{figure}

 The ANGST observations from which the CMD SFHs were generated are publicly available. To gauge the level of incompleteness and in hopes of including additional bands in our analysis, we create faux integrated ACS photometry by summing the fluxes of the stars in the footprint of several galaxies. We plot these points on the SED to get a rough estimate of the agreement between the integrated photometry and the CMD footprint as shown in Figure ~\ref{fig:gst_1}. For this initial test we only used galaxies that weren't mosaicked. We created mock bands by simply summing over the star files. We compared these to the broadband photometry, as shown in red in the figure, as well as the spectra generated from the CMD SFHs, as shown in black. In Figure ~\ref{fig:gst_1}, we show four of the more massive galaxies, where we expect the greatest signal to noise. In this rudimentary test, our findings were that 1) there was enough agreement between our summed bands, the generated spectra, and the broadband photometry to validate the data we were already working with, but 2) this rudimentary test showed offsets in the summed bands as compared to the broadband photometry that recommended against adding these bands for this study.

\section{SEDs and SFHs of individual galaxies}
\label{appendix:individual_sfhs}
The 36 ANGST galaxies selected for this study represent a diverse set of morphologies, environments, and SFHs. We show individual SEDs and SFHs for each galaxy in the following three figures. Each row represents one galaxy, with the SED on the left and the SFH on the right. The SED shows the spectrum inferred from the CMD SFH in black. Overlaid are the mock photometry in blue and the observed broadband photometry in red. In the SFH panels we show the CMD SFH in black, and then the SFHs with uncertainties corresponding to the mock and observed photometries in blue and red respectively. Galaxies are ordered by increasing  sSFR. 

\begin{figure}[H]
    \centering
    
    \includegraphics[width = 0.88\textwidth]{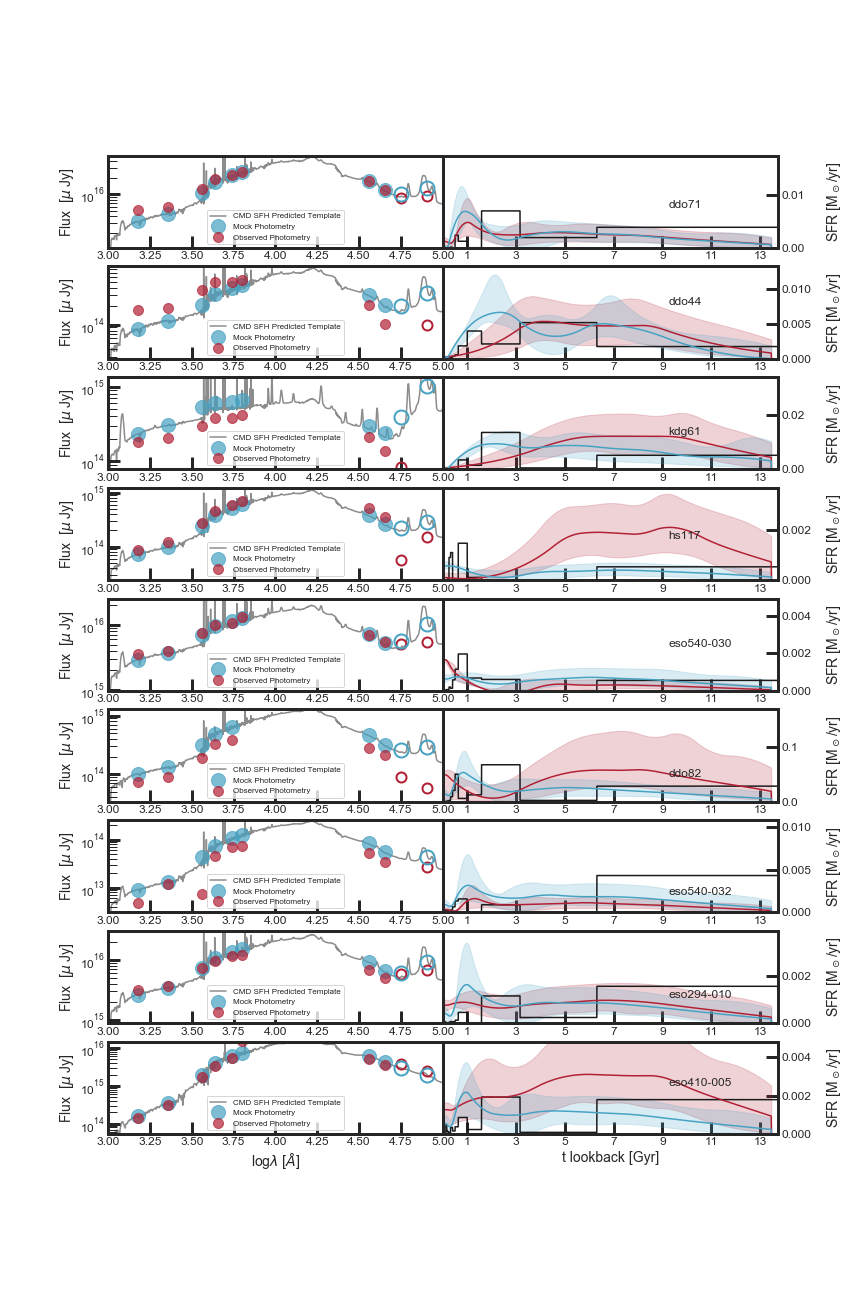}

    \caption{Galaxy SEDs and SFHs ordered in increasing sSFR. The SED shows the spectrum inferred from the CMD SFH in black. The mock photometry are in blue and the observed broadband photometry in red. SFH panels show the CMD SFH in black, and SFHs with corresponding to the mock and observed photometries in blue and red respectively.}
    \label{fig:sed_sfh0}
\end{figure}

\begin{figure}[H]
    \centering
    
    \includegraphics[width = 0.88\textwidth]{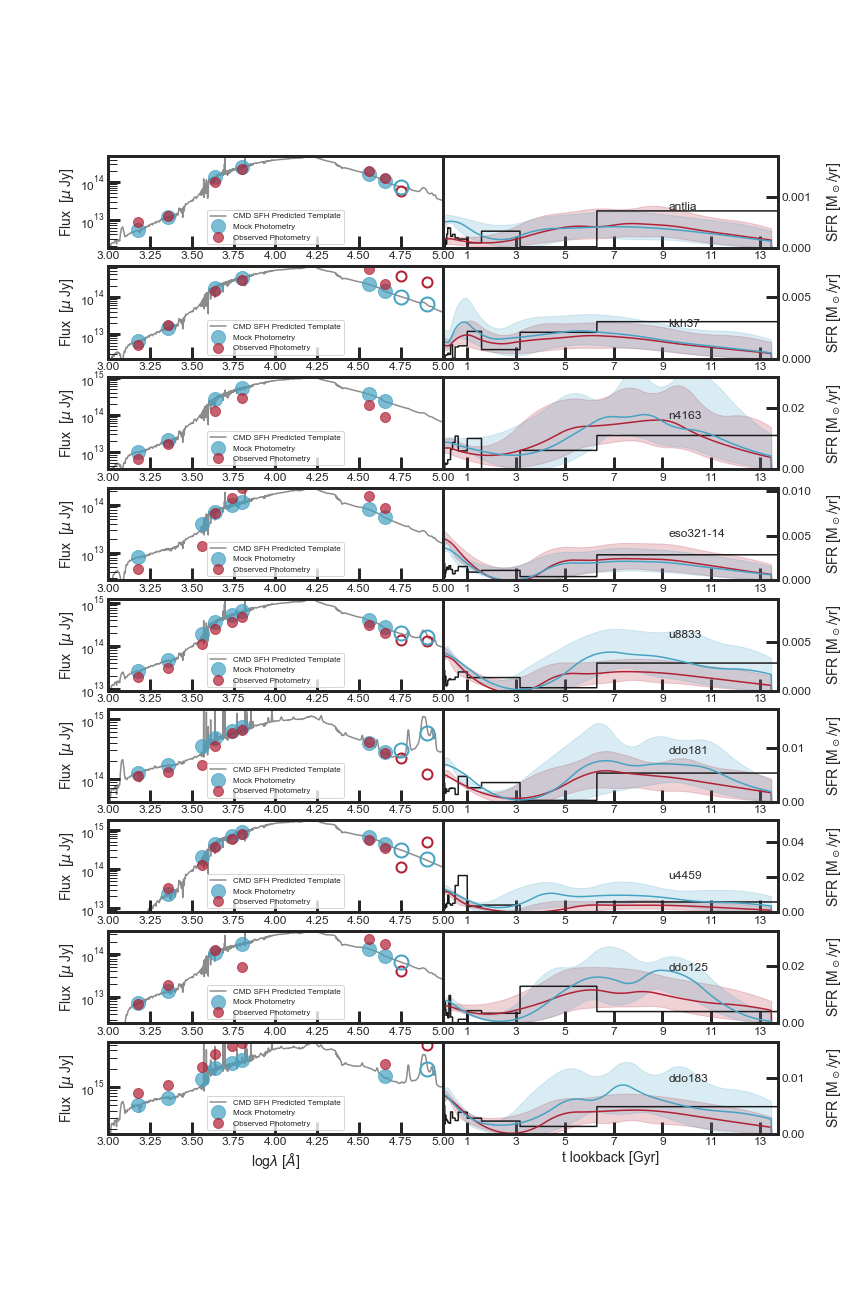}

    \caption{Galaxy SEDs and SFHs ordered in increasing sSFR. The SED shows the spectrum inferred from the CMD SFH in black. The mock photometry are in blue and the observed broadband photometry in red. SFH panels show the CMD SFH in black, and SFHs with corresponding to the mock and observed photometries in blue and red respectively.}
    \label{fig:sed_sfh1}
\end{figure}

\begin{figure}[H]
    \centering
    
    \includegraphics[ width = 0.88\textwidth]{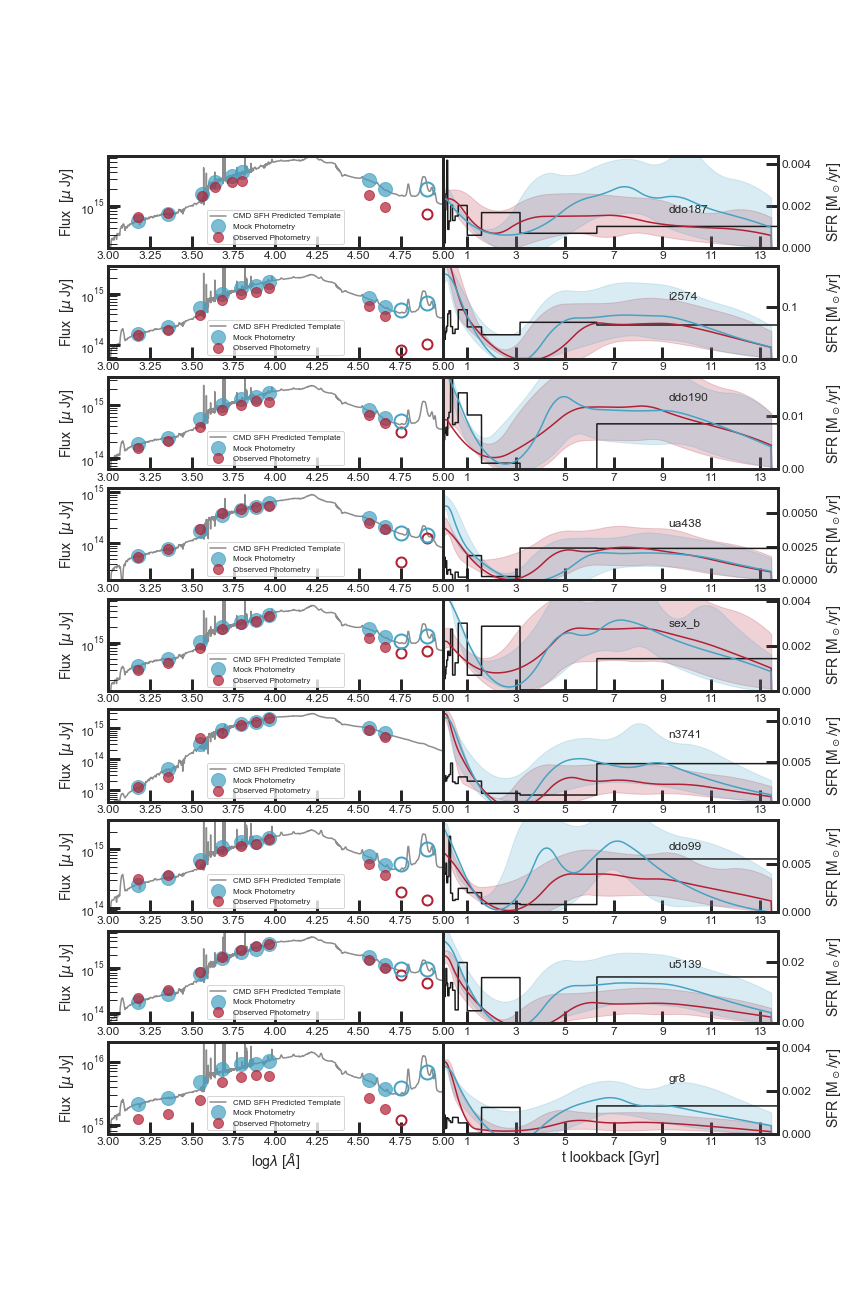}

    \caption{Galaxy SEDs and SFHs ordered in increasing sSFR. The SED shows the spectrum inferred from the CMD SFH in black. The mock photometry are in blue and the observed broadband photometry in red. SFH panels show the CMD SFH in black, and SFHs with corresponding to the mock and observed photometries in blue and red respectively.}
    \label{fig:sed_sfh2}
\end{figure}

\begin{figure}[H]
    \centering
    
    \includegraphics[width = 0.88\textwidth]{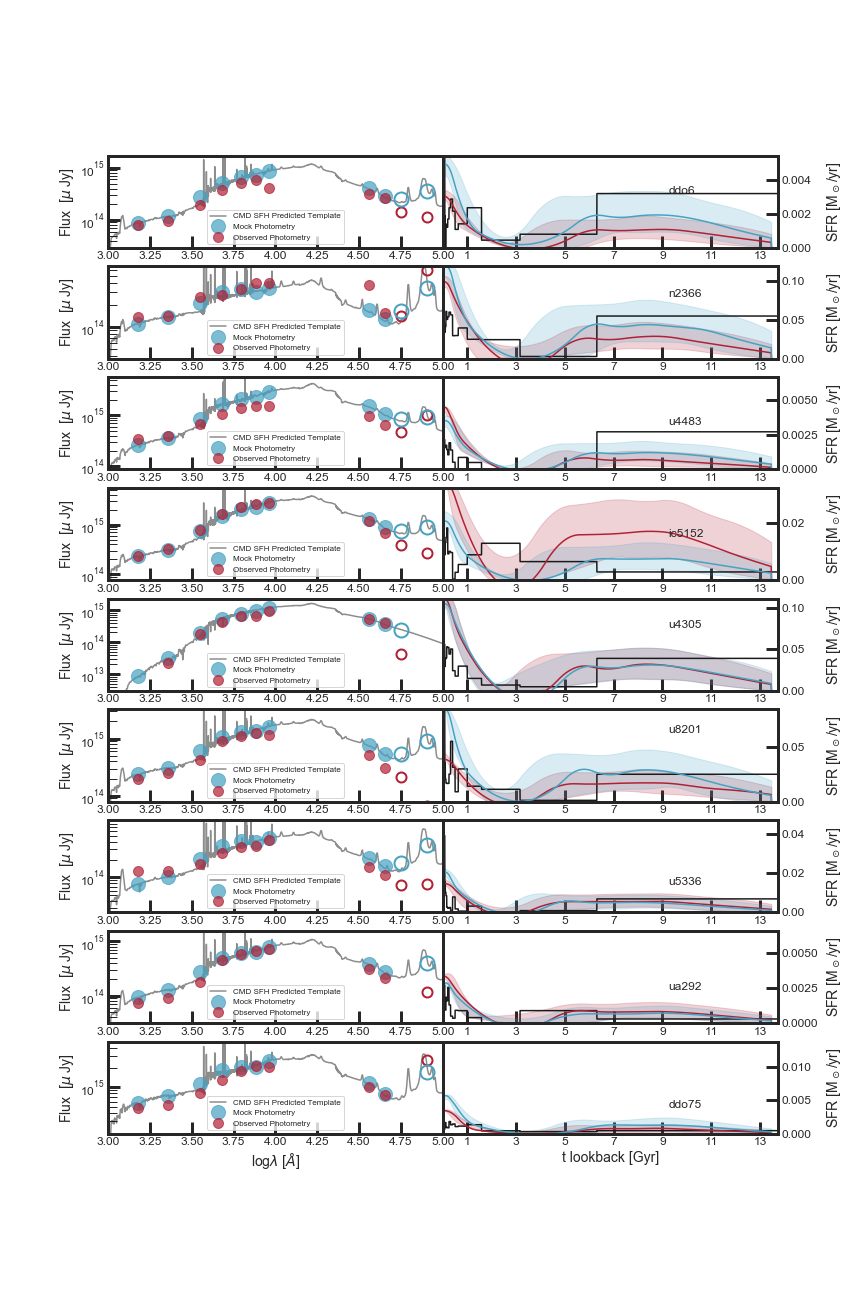}

    \caption{Galaxy SEDs and SFHs ordered in increasing sSFR. The SED shows the spectrum inferred from the CMD SFH in black. The mock photometry are in blue and the observed broadband photometry in red. SFH panels show the CMD SFH in black, and SFHs with corresponding to the mock and observed photometries in blue and red respectively.}
    \label{fig:sed_sfh3}
\end{figure}

\clearpage

\bibstyle{apj}
\bibliography{library}


\listofchanges

\end{document}